\pdfoutput=1

\documentclass[final,1p,authoryear]{elsarticle}

\usepackage{float}
\usepackage{amssymb}

\usepackage{amsmath,amsfonts}
\usepackage{todonotes}
\usepackage{booktabs}
\usepackage{siunitx}
\usepackage{subfig}
\usepackage{hyphenat}
\usepackage{xurl}
\usepackage[breaklinks=true]{hyperref}
\usepackage{textpos}

\usepackage{verbatim}
\usepackage{multirow} 
\usepackage{graphicx}

\usepackage{todonotes}

\usepackage{subcaption}
\usepackage{xcolor}

\usepackage{adjustbox}

\usepackage{longtable}
\usepackage{rotating}

\usepackage{booktabs}

\makeatletter
\def\ps@pprintTitle{   \let\@oddhead\@empty
   \let\@evenhead\@empty
   \def\@oddfoot{\reset@font\hfil\thepage\hfil}
   \let\@evenfoot\@oddfoot
}
\makeatother

\graphicspath{{./img/}}

\begin{document}

\begin{frontmatter}

\title{Computational Analysis of Speech Clarity Predicts Audience Engagement in TED Talks}

\author[label1]{Roni Segal\fnref{equal}}
\author[label1]{Matan Lary\fnref{equal}}
\author[label2]{Ralf Schmaelzle}
\author[label1]{Yossi Ben-Zion}\ead{benzioy@biu.ac.il}

\fntext[equal]{These authors contributed equally to this work.}

\address[label1]{Department of Physics, Bar Ilan University, 
Ramat Gan 52900, Israel}
\address[label2]{Department of Communication, Michigan State University, 
East Lansing 48824, MI, USA}

\begin{abstract}
What makes a public talk resonate with large audiences? While prior research has emphasized speaker delivery or topic novelty, we reasoned that a core driver of engagement is linguistic clarity. This aligns with theories of processing fluency and cognitive load, which posit that audiences reward speakers who present complex ideas accessibly.

We leveraged artificial intelligence to analyze 1,239 TED Talk transcripts (2006--2013), supplemented by a later-phase longitudinal sample. Each transcript was evaluated across 50 independent large language model runs on two dimensions, clarity of explanation and structural organization, and linked to YouTube engagement metrics (likes and views).

Clarity emerged as the strongest predictor of audience responses ($\beta = .339$ for likes; $\beta = .314$ for views), contributing substantial incremental variance ($\Delta R^{2} \approx .095$) beyond duration, topic, and scientific status. The full model explained 29\% of variance in likes and 22.5\% in views. This effect was domain-general, remaining invariant across content categories and between scientific and non-scientific talks. Notably, clarity outperformed traditional readability metrics, indicating that discourse coherence predicts engagement more powerfully than surface-level linguistic simplicity. Longitudinal analyses further revealed standardization within TED, characterized by increasing clarity and reduced variability over time.

Theoretically, these results support processing fluency accounts: clearer communication reduces cognitive friction and elicits more positive evaluative responses. Practically, transcript-based clarity represents a scalable and trainable strategy for improving public discourse. By demonstrating that language models can reliably capture latent communicative qualities, this study paves the way for feedback systems in education, science communication, and public speaking.
\end{abstract}


\end{frontmatter}

\section{Introduction}
\label{sec:intro}

Why do some talks engage millions of people while other ideas, perhaps equally substantive and important, are ignored? This disparity in audience engagement has massive consequences for how knowledge spreads in society. Past research has rightfully looked at variables such as topic interest, emotional arousal, or speaker charisma to explain why some content succeeds online \citep{berger2012makes, sugimoto2013scholars}. However, the present study focuses on a much more straightforward, and perhaps therefore overlooked, message-level factor: linguistic clarity. 

We view clarity essentially as a measure of signal fidelity. If the signal of an idea or message is clear, well-structured, and easy to process, it should theoretically lead to greater audience engagement, especially when scaled to a global digital audience. To test this, we leveraged artificial intelligence (AI) to evaluate the explanatory clarity of TED Talks, examining how this core communicative property drives large-scale engagement on YouTube.

To investigate how clarity impacts audience engagement, this paper proceeds as follows. First, we establish a theoretical framework explaining why message-level clarity systematically impacts aggregate audience behavior. We then introduce an AI-driven methodology used to holistically evaluate transcripts, providing a scalable way to quantify communicative quality. Finally, we present an empirical analysis testing whether clarity serves as a domain-general predictor of engagement and examine how these communicative standards have evolved over time within the TED ecosystem.

\subsection{Why Clarity Matters: From Message Property to Collective Engagement}

Mass communication, public speaking, and many aspects of classroom education are fundamentally a one-to-many communication activity: a single message is broadcast to multiple recipients. While any mass audience is characterized by vast individual differences, varying levels of prior knowledge, interest, motivation, and cognitive ability, the linguistic clarity of the message remains a constant, stimulus-sided factor. From the perspective of text comprehension \citep{mcnamara2013reading, kintsch1998comprehension}, clearer messages convey information with higher fidelity and less signal loss. Because a clear message successfully navigates the constraints of human information processing, it should affect diverse audience members in systematically similar ways, yielding downstream consequences for engagement \citep{schmaelzle2022theory}.

The theoretical mechanism linking message clarity to positive audience evaluation is rooted in the concept of processing fluency \citep{reber2004processing}. The processing fluency framework posits that information that is easy to process is not only better understood but also elicits a more positive affective response. In the context of mass communication or science journalism, this manifests as a "simple-writing heuristic," wherein audiences inherently reward online texts that minimize cognitive friction \citep{shulman2024reading, bullock2021narratives}.

Conversely, a lack of clarity is cognitively costly and subjectively aversive. Decades of research on cognitive load \citep{sweller1988cognitive} and recent meta-analytic evidence confirm that the mental effort required to decode complex or disorganized information is inherently unpleasant \citep{david2024unpleasant}. This friction acts as a barrier to engagement, a phenomenon vividly illustrated by the negative "consequences of erudite vernacular," wherein speakers or writers who use unnecessarily complex language are often judged more harshly \citep{oppenheimer2006consequences}. Optimal engagement occurs when the challenge of the content is matched by the accessibility of its delivery, reducing mental friction and facilitating a state of communicative flow \citep{csikszentmihalyi91flow}.
While clarity is clearly not the only factor driving popularity \citep{scholz2017neural}, it is a foundational one. At the individual level, the metacognitive reward of processing a clear explanation might seem small. However, when a single message is viewed by thousands or millions of individuals on platforms like YouTube, even subtle cognitive efficiencies aggregate into stable, macro-level effects.

From this perspective, YouTube views and likes are not mere vanity metrics; they are objective, revealed preferences that capture the behavioral consequences of collective audience processing. A "view" represents the decision to consume or sustain attention, while a "like" functions as a quantifiable, evaluative behavioral statement, a form of social currency distributed when a viewer feels rewarded by the content \citep{sherman2018brain}. Predicting these objective markers is fundamentally more robust than relying on self-reported motivations, especially when the motivational signal is only weak, fleeting, and likely implicit in nature.

It is worth noting that prior research has attempted to quantify text accessibility using readability formulas (e.g., \citep{flesch1948new}). While foundational, these metrics relied heavily on calculating word length and sentence boundaries rather than capturing the holistic, semantic, and structural coherence of a spoken explanation, leading to persistent concerns regarding their validity for evaluating actual communicative quality.

\subsection{AI-Based Evaluation and the TED Talk Context}

Recent advances in large language models (LLMs) provide us with a transformative tool. AI models can now holistically evaluate the semantic architecture, logical flow, and explanatory clarity of extended texts. In particular, prior work by \citet{zion2025ai} successfully utilized LLMs to evaluate the communicative quality of public speaking and university physics lectures. Analyzing  32 physics courses (1,222 lecture hours) at Bar-Ilan University, they found that AI-based transcript evaluations of clarity and structure not only correlated strongly with student perceptions but also doubled the explanatory power of regression models (from 18.9\% to 38.3\%), outperforming traditional predictors such as course grades and class size. Relatedly, \citet{schmalzle2025art} demonstrated that AI-derived evaluations of scientific talks strongly align with human judgments. These studies  demonstrate that transcript-based approaches  offer scalable, interpretable, and replicable means of assessing communication quality and link these measures to outcomes.

To test the impact of clarity at scale, we apply this methodology to TED Talks. The TED platform represents the ultimate testbed for science communication and public speaking, characterized by its global reach, academic credibility, and standardized format. 

Previous computational research on TED Talks has provided valuable insights, but it has largely focused on extracting peripheral cues. For instance, studies have engaged in the counting of nonverbal gestures \citep{cascio2024talking} or relied on sentiment analysis to track emotional density \citep{fischer2024affect}. By focusing heavily on the affective and visual layers of the presentations, the core communicative channel, the linguistic clarity of the talk itself, has remained less examined.

\subsection{The Professionalization of TED and the Research Gap}

Despite the wealth of data on online popularity, a significant empirical gap remains. Most transcript-based and metadata-driven studies have focused on nonverbal cues, video duration, or topic tags rather than the structural and linguistic clarity of the message. Without a scalable, transcript-based measure of communicative quality, it has been difficult to determine whether linguistic accessibility is a unique, independent predictor of engagement, or merely a byproduct of topic selection.

Furthermore, while TED is widely cited as the benchmark for "ideas worth spreading," little empirical work has examined the platform's historical evolution. It is widely acknowledged that TED has undergone significant institutional professionalization over the past two decades \citep{ludewig2017ted}, but it remains unknown whether this has led to a measurable standardization of linguistic clarity over time.

Institutional practices reinforce this structure. Speakers are heavily coached to identify a clear “throughline” and shape their presentations around a concise central idea \citep{anderson2016ted}. Taken together, these processes suggest that TED Talks increasingly reflect deliberate communicative design. Yet, despite this widely acknowledged professionalization, little empirical work has examined whether it is reflected in measurable patterns of communicative clarity over time.

\subsection{The Current Study and Hypotheses}

The present study addresses these gaps by combining large-scale transcript analysis with AI-based discourse evaluation. Using repeated large language model assessments of TED Talk transcripts, we introduce a scalable measure of linguistic clarity and examine its relationship to audience engagement metrics such as likes and views, as well as its evolution across different stages in the development of the TED platform.

\begin{figure}[ht]
    \centering
    \includegraphics[width=0.90\textwidth]{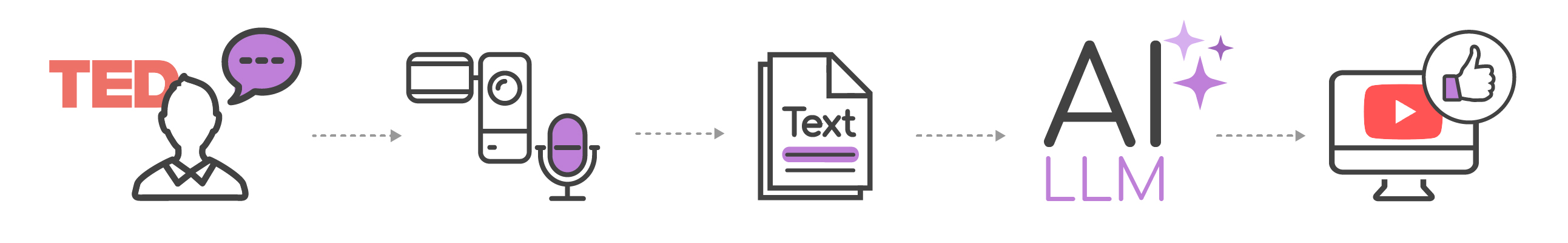}
    \caption{Overview of the AI-based transcript evaluation pipeline for TED Talks. A large corpus of curated transcripts is evaluated using repeated large language model (LLM) assessments, which are subsequently linked to large-scale audience engagement metrics on YouTube, enabling high-resolution inference of linguistic predictors of engagement.}
    \label{fig:ai_process}
\end{figure}

We build on a large corpus of 1,239 TED Talks that is examined via structured large language model (LLM) evaluations and linked to YouTube engagement metrics. Specifically, each TED Talk is converted into a curated transcript and independently evaluated across 50 LLM runs using a domain-adapted prompt to generate statistically robust estimates of linguistic clarity and structure. These AI-derived scores are then systematically linked to large-scale behavioral engagement indicators (likes and views). Hierarchical regression models and extensive robustness analyses are employed to isolate the unique contribution of linguistic clarity beyond baseline temporal, topical, and scientific factors. The overall AI-based evaluation pipeline employed in this study is illustrated in Figure~\ref{fig:ai_process}.

This framework enables a systematic, high-resolution examination of how linguistic clarity operates as a central and domain-general predictor of audience engagement, motivating the research questions addressed in the present study.

Based on the theoretical account outlined above, we proposed the following hypotheses and research questions:
\begin{itemize}
  
    \item Hypothesis 1 (H1):  Linguistic clarity (comprising explanatory clarity and structural organization) serves as a primary driver of audience resonance in digital environments.
    
    Specifically:
    H1a (Main Effect): AI-derived clarity scores from TED Talk transcripts will predict audience engagement (view counts and likes).
    
    H1b (Incremental Validity): Clarity scores will explain variance in engagement metrics beyond baseline factors, including duration, trends, and topical categories.
 
    \item Research Question 1 (RQ1): To what extent does the predictive relationship between clarity and audience engagement depend on the nature of the content (scientific vs. non-scientific talks)?

    \item Research Question 2 (RQ2): How has the distribution of clarity evolved across the TED platform's early vs. consolidated phase and what do longitudinal trends reveal about the professionalization and standardization of the TED genre?
\end{itemize}

\section{Methodology}
\label{sec:Methodology}

The present study aims to examine whether and how transcript-based communicative qualities, as assessed by large language models (LLMs), can explain and predict large-scale audience engagement with public talks. Specifically, we investigate the extent to which AI-derived evaluations of explanatory clarity and structural organization from TED Talk transcripts predict behavioral engagement on YouTube, including view counts and likes, and whether these linguistic attributes provide explanatory power beyond established baseline predictors. In addition, we examine whether these relationships differ between scientific and non-scientific talks, and how the distribution of communicative quality evolves across different stages in TED’s institutional development.

To address these research questions, we developed an AI-based analytical framework that integrates transcript-level discourse evaluation with platform-scale behavioral metrics. This framework enables systematic, scalable, and theory-driven measurement of latent communicative attributes and their association with audience engagement in naturalistic public communication settings.

Our study adopts a two-phase longitudinal design, distinguishing between an early formative phase and a later mature phase of the platform, in line with documented changes in TED’s institutional practices and production standards over time.

The primary analysis focuses on the foundational period of TED (2006–2013), during which the platform underwent rapid expansion and substantial refinement of its editorial, rhetorical and production practices. This period is characterized by pronounced heterogeneity in presentation styles and communicative quality, making it particularly suitable for examining how transcript-based clarity and structure relate to audience engagement and how these relationships evolve over time.

To assess whether these patterns persist under conditions of advanced professionalization and reduced stylistic diversity, we additionally analyze a later phase represented by the years 2017 and 2019. These years were selected as representative snapshots of TED’s mature stage, capturing a period in which production standards, speaker coaching, and genre conventions were already well established, while avoiding potential confounds introduced by the COVID-19 pandemic and its effects on content production, dissemination, and audience behavior. Together, these two time windows provide complementary perspectives on early-stage development and later-stage stabilization, enabling an examination of both longitudinal change and potential ceiling effects in AI-based discourse evaluation.

Across both phases, each TED Talk is converted into a curated transcript and independently evaluated using repeated large language model (LLM) assessments, yielding statistically robust estimates of explanatory clarity and structural organization. These AI-derived discourse measures are subsequently linked to large-scale audience engagement indicators, including log-transformed view counts and likes, alongside temporal exposure measures derived from Google Trends. Hierarchical regression models and extensive robustness analyses are employed to isolate the unique contribution of linguistic clarity and structure beyond baseline temporal, topical, and disciplinary factors.

\subsection{YouTube as the Primary Platform for TED Talks}

Since 2007, TED has systematically distributed its talks through YouTube, which has gradually become the dominant and most legitimate viewing arena for TED content. As YouTube grew into one of the world’s largest video-sharing platforms, it also became the de facto channel through which global audiences access TED’s educational and scientific material. This centrality of YouTube justifies its use as the primary data source for the present study.

Key metadata for each video (including the number of views, publication date, duration, and number of likes) were retrieved via the YouTube Data API. These variables served as standardized and transparent indicators of audience engagement, widely used in previous communication and media studies.

Beyond these engagement metrics, textual data also played a crucial role in the present analysis. TED provides human-generated transcripts for nearly all talks, which are prepared manually by TED’s editorial team or by volunteers in the TED Translators program. These transcripts represent a far more reliable textual source than the automatically generated captions available on YouTube. In many cases, YouTube defaults to an auto-generated English transcript, which is often incomplete or inaccurate. In contrast, TED’s human-edited versions preserve linguistic nuance, correct punctuation, and contextual meaning, minimizing transcription errors that could distort textual analysis.

Because the present study focuses on AI-based measures of clarity and structure, both of which are sensitive to sentence boundaries and phrasing, the use of TED’s human-generated transcripts was essential for ensuring textual accuracy and interpretive validity.

\subsubsection{Temporal Distribution of TED Talks}

The dataset spans TED Talks published between December 2006 and December 2013, corresponding to the period when TED systematically expanded its online presence through YouTube.  
Table~\ref{tab:year_distribution_combined} presents the yearly distribution of TED Talks across the study phases, including both the early sample (2006--2013) and the later mature sample (2017 and 2019).

As shown for the early phase, the number of uploaded talks grew steadily throughout the examined period, reflecting TED’s gradual expansion of its online presence.
The dataset includes only a single talk from 2006, corresponding to TED’s initial experimental uploads, followed by a consistent increase in the number of talks each year, peaking in 2013.
This gradual growth mirrors the overall rise in global search interest for TED observed in the Google Trends analysis (Figure~\ref{fig:googletrends}).

\begin{table}[htbp]
\centering
\caption{Yearly distribution of TED Talks across the early and late study phases.}
\label{tab:year_distribution_combined}
\renewcommand{\arraystretch}{1.1}
\begin{tabular}{llcc}
\hline
\textbf{Phase} & \textbf{Year} & \textbf{Frequency} & \textbf{Percent} \\
\hline
\multirow{9}{*}{Early phase}
& 2006 & 1   & 0.1 \\
& 2007 & 107 & 8.6 \\
& 2008 & 146 & 11.8 \\
& 2009 & 176 & 14.2 \\
& 2010 & 190 & 15.3 \\
& 2011 & 196 & 15.8 \\
& 2012 & 200 & 16.1 \\
& 2013 & 223 & 18.0 \\
& \textbf{Total} & \textbf{1,239} & \textbf{100.0} \\
\hline
\multirow{3}{*}{Late phase}
& 2017 & 198 & 41.2 \\
& 2019 & 283 & 58.8 \\
& \textbf{Total} & \textbf{481} & \textbf{100.0} \\
\hline
\end{tabular}
\end{table}

In addition to the foundational period (2006--2013), a later mature phase was analyzed to examine communicative patterns under conditions of advanced institutional standardization. This late phase includes TED Talks published in 2017 and 2019, selected as representative years following the platform’s consolidation while avoiding potential distortions associated with the COVID-19 pandemic. These years were not selected to form a continuous time series but rather as representative snapshots of TED’s mature stage, enabling comparison with the more heterogeneous early period. As shown in Table~\ref{tab:year_distribution_combined}, the late-phase sample consisted of 198 talks from 2017 and 283 talks from 2019.

\subsubsection{Distribution of Engagement Metrics}

The distributions of YouTube engagement metrics were highly right-skewed: most TED Talks received relatively few views and likes, whereas a small number accumulated exceptionally high engagement. Such heavy-tailed distributions are common in online media, where attention tends to concentrate on a limited number of highly popular items. Because these variables spanned several orders of magnitude, the dependent variables were log-transformed (\textit{log(Views)} and \textit{log(Likes)}). This transformation reduced skewness, compressed extreme values, and produced distributions that were approximately normal, thereby improving interpretability and better satisfying the assumptions of subsequent linear regression analyses.

\begin{figure}[htbp]
    \centering
    \subfloat[Views\label{fig:log_views_histogram}]{
        \includegraphics[width=0.48\textwidth]{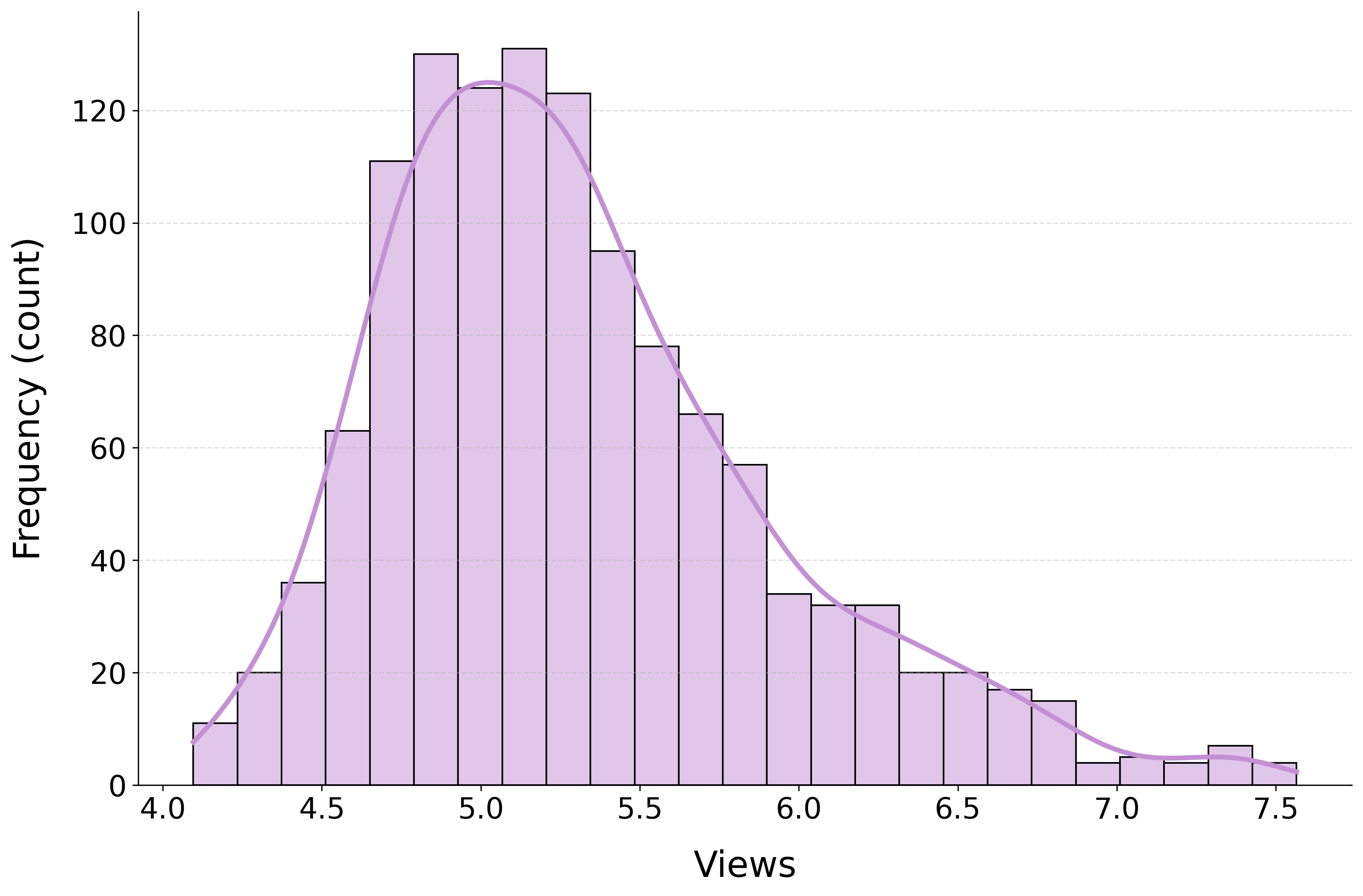}
    }
    \hfill
    \subfloat[Likes\label{fig:log_Likes_histogram}]{
        \includegraphics[width=0.48\textwidth]{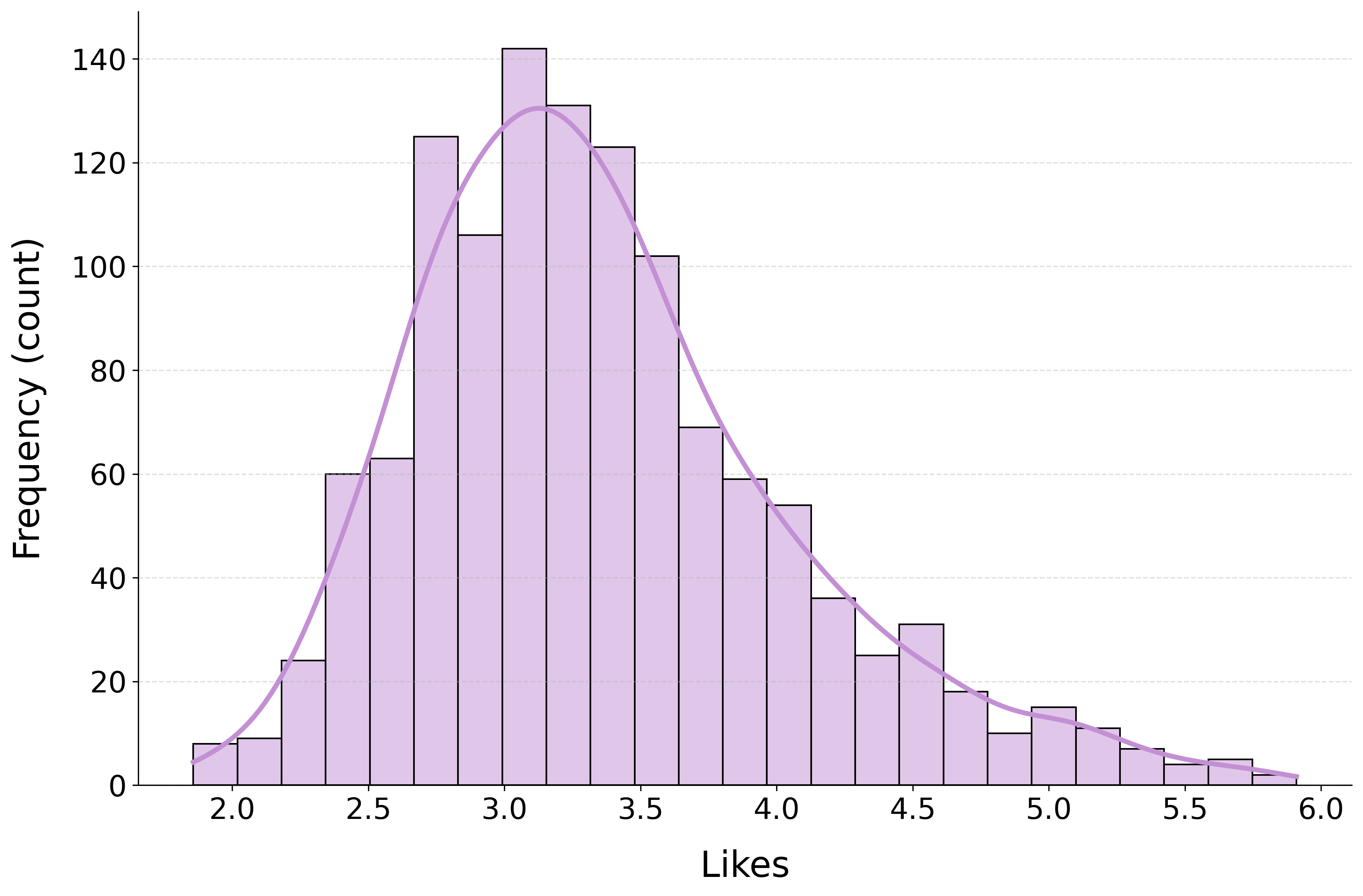}
    }
    \caption{Histograms of the log-transformed TED Talk engagement metrics.}
    \label{fig:log_engagement_histograms}
\end{figure}

Figure~\ref{fig:log_engagement_histograms} presents the histograms of the log-transformed distributions of views and likes. From this point onward, the terms \textit{Views} and \textit{Likes} refer to their log-transformed values for simplicity.

\subsubsection{Descriptive Statistics of YouTube Engagement Metrics}

Table~\ref{tab:youtube_descriptives} presents descriptive statistics for the key quantitative variables extracted from the YouTube Data API, including video duration, logarithmically transformed view counts, and like counts. The dataset comprises 1,239 TED Talks. As shown, the talks varied substantially in length (ranging from approximately 2 to 35 minutes), and both engagement metrics (\textit{Views} and \textit{Likes}) exhibited wide variability even after logarithmic transformation, reflecting the heterogeneity of audience attention typical of large-scale online media datasets.

\begin{table}[htbp]
\centering
\caption{Descriptive statistics of YouTube engagement variables (N = 1,239).}
\label{tab:youtube_descriptives}
\begin{tabular}{lccccc}
\hline
\textbf{Variable} & \textbf{N} & \textbf{Minimum} & \textbf{Maximum} & \textbf{Mean} & \textbf{Std. Deviation} \\
\hline
Duration (s) & 1,239 & 136 & 2,111 & 871.64 & 352.07 \\
Views & 1,239 & 4.10 & 7.56 & 5.31 & 0.62 \\
Likes & 1,239 & 1.86 & 5.91 & 3.36 & 0.69 \\
\hline
\end{tabular}
\end{table}

\subsubsection{Excluded Variables and Rationale for Omission}

Some variables retrieved from the YouTube Data API were excluded from the final analyses for conceptual and methodological reasons. These included the number of translation languages, the number of audience comments, and linguistic pace indicators such as words per second.

The number of translation languages was removed because it likely reflects an outcome of audience engagement rather than a predictor. Although the number of comments was highly correlated with views and likes, we excluded it since it mainly reflects audience activity following popularity rather than an independent factor. Finally, speech rate and word count were excluded due to high collinearity with video duration, which was chosen because it showed the strongest correlations with the main study variables.

\subsection{Google Trends as a Temporal Context Variable}
\label{subsec:google_trends}

To control for fluctuations in external interest over time, we incorporated data from Google Trends, a public analytics tool that reports the relative popularity of search queries on a normalized scale from 0 to 100. This index approximates the proportion of searches for a given query relative to all Google searches within a specific time frame and region, providing a standardized measure of public attention.

A distinctive feature of Google Trends is its topic search function, which aggregates semantically related queries across languages under a single conceptual entity. For example, searching for the topic “TED” encompasses related queries such as “TED Talks”, “TEDx”, or “TED conference”. This method yields a more accurate and language-independent measure of global interest in the TED phenomenon itself.

In this study, we used Google Trends to quantify the overall public interest in TED during each talk’s release period, independently of YouTube activity. This variable served as a temporal context factor, allowing us to account for shifts in TED’s baseline popularity when analyzing engagement metrics and AI-derived clarity and structure scores.

Hereafter, this variable is referred to as the \textit{TED Google Trends index} and denoted as \textit{TED\_TrendIndex} in all statistical analyses.

The analysis covered the period from December 25, 2006, corresponding to the release of the first TED Talk included in the dataset, through December 23, 2013, the release date of the final talk analyzed. These boundaries were selected to align the Google Trends data with the actual temporal range of the videos in our corpus.

It is important to note that Google Trends does not report absolute search volumes, but rather a normalized measure known as Relative Search Volume (RSV). For a given topic $k$, region $r$, and time window $T$, the reported value at time $t \in T$ reflects the proportion of searches for that topic relative to all Google searches at that time, scaled to the peak interest observed within the period:

\begin{equation}
RSV(k,r,t) = 100 \times 
\frac{\left( \frac{n(k,r,t)}{N(r,t)} \right)}
{\max_{\tau \in T} \left( \frac{n(k,r,\tau)}{N(r,\tau)} \right)} .
\end{equation}

This normalization ensures that the maximum relative interest within the analyzed time window is set to 100, allowing comparisons of temporal trends in public attention independent of absolute search volume.

To visualize these temporal fluctuations in public attention, the corresponding Google Trends data are presented in Appendix~\ref{app:google_trends}.

\subsection{AI-Based Evaluation of Clarity and Structure}
\label{subsec:ai_clarity}

This study builds upon the methodological framework developed in our previous work on AI-based teaching evaluations (\cite{zion2025ai}), where linguistic indicators of \textit{clarity} and \textit{structure} were extracted from transcripts of university physics lectures. In that study, large language models (LLMs) showed strong correlations with human evaluations, suggesting that transcript-based AI assessments can serve as reliable proxies for perceived teaching quality. In addition, repeated evaluations of the same transcripts produced highly consistent results: for each lecture, the distribution of AI-generated scores closely followed a near-normal distribution centered around the mean.
This convergence indicates that the model’s outputs were not random or unstable but instead reproducible across runs, reflecting statistical reliability and internal coherence in its judgments.

In the present work, this approach was extended to TED Talk transcripts sourced from YouTube, in order to examine whether the same linguistic dimensions predict audience engagement metrics such as views and likes.

To ensure methodological continuity, the same evaluation framework from \cite{zion2025ai} was adopted, with minor adaptations to fit the TED context.
Two prompt versions were used: the original university-level prompt (included in Appendix~\ref{app:prompt}) and an adapted TED-specific version.
The original prompt was applied only using ChatGPT (GPT-4o), whereas the TED-specific prompt was tested with ChatGPT, Gemini and Claude to examine cross-model consistency.
The main analyses presented in this paper are based on the ChatGPT (TED version) runs, while comparative results between prompt and model types are reported for validation and robustness assessment.
The model was prompted to assess each transcript based on two dimensions: \textit{clarity of explanation} and \textit{lecture structure and logical flow}. The adapted prompt used in this study was as follows:

\begin{quote}
\textbf{Prompt (TED version)}\\
You will serve as an expert in evaluating \textbf{TED lectures}.\\
Your task is to assess the quality of a TED lecture based on the following two criteria:\\[4pt]
\textit{Clarity of Explanation (1–10)}\\
\textit{Lecture Structure and Logical Flow (1–10)}\\[4pt]
Evaluate based on a transcript of a lecture where only the lecturer's speech is transcribed.\\
Provide a score between 1 and 10 for each criterion, without further explanation.\\
Your response should be in the format: \texttt{X,X} (e.g., \texttt{8,9})\\[4pt]
\{transcript\}
\end{quote}

Each transcript was independently evaluated across 50 runs, and the mean score for each criterion (clarity and structure) was computed per talk, resulting in a single pair of averaged values per talk.
For brevity, throughout the remainder of this paper, \textit{Clarity of Explanation} is referred to simply as \textit{Clarity}, and \textit{Lecture Structure and Logical Flow} as \textit{Structure}.

Furthermore, the use of AI-based linguistic assessment is supported by recent empirical evidence.
In a related study, \cite{schmalzle2025art} applied large language models to evaluate over 100 scientific talks using similar clarity-oriented prompts.
Their findings demonstrated strong alignment between AI-generated evaluations and human ratings, even when the AI model was provided with only the opening excerpts of each talk (less than 10\% of the transcript).
These results support the validity of using LLMs to capture genuine communicative qualities such as clarity and structure lending external legitimacy to the present methodology.

\subsection{Scientific Classification}
\label{subsec:Sci_class}

Following \cite{fischer2024affect}, who assessed the scientific nature of TED Talks based on TED-assigned tags, we adopted a conceptually similar yet methodologically distinct approach. Instead of relying on metadata, we evaluated the scientific character of each talk using the transcript itself.

The classification format described below includes both the scientific and topical components. While both were generated within the same prompt, the present subsection focuses on the scientific flag, and the following subsection elaborates on the topical categories.

The model was prompted with the following format:

\begin{quote}
You will serve as an expert in evaluating TED lectures.
You are classifying a TED Talk transcript.

\textbf{Task A — Scientific flag:}  
\\ - Output 1 if the talk is primarily scientific, meaning the content is based on scientific research, empirical evidence, or established scientific concepts (e.g., experiments, data, peer-reviewed findings).
\\ - Output 0 if the talk mainly uses stories, metaphors, inspiration, or philosophy without focusing on the scientific method or evidence. 

\textbf{Task B — Category (choose exactly one):}  
Health, Cosmos, Mind, Environment, Tech, Society, Entertainment 

Evaluate based on a transcript of a lecture where only the lecturer's speech is transcribed. 
Your response should be in the format: \texttt{S,CATEGORY} (e.g., \texttt{1,Tech})  
where $S \in \{0,1\}$ and \texttt{CATEGORY} $\in$ \{Health, Cosmos, Mind, Environment, Tech, Society, Entertainment\}.
\end{quote}

Unlike prior tag-based approaches, this transcript-based method leverages contextual understanding.
Instead of relying on isolated words that may carry different meanings in different settings, the model interprets each statement in relation to the surrounding discourse, capturing the full communicative intent of the talk.
This allows for a more accurate and conceptually grounded distinction between scientific and non-scientific content.

Each transcript was evaluated fifteen times, producing binary scientific scores (0 or 1) for each repetition.  
The mean of these values was calculated to obtain a continuous scientificness score for each talk.  
Talks with an average score greater than 0.5 were classified as scientific, and those with a mean score below 0.5 were classified as non-scientific.

A histogram of these averaged scores (see Figure~\ref{fig:scientific_hist} in Appendix~\ref{app:scientific_hist}) illustrates their distribution, showing that most values clustered around 0 and 1 (meaning that all runs yielded the same classification). Out of all evaluated talks, 86.4\% received a mean score of exactly 0 or 1, while only 13.6\% fell between these extremes, and merely 3.5\% were in the mid-range (0.3–0.7). This confirms that the binary classification was largely unambiguous and internally consistent across repeated evaluations.

This bimodal pattern further supports the reliability of the classification procedure, indicating that the model consistently converged to stable judgments across repeated evaluations.

The topic component of this classification is described in detail in the following subsection.

\subsection{Topic Classification and Reliability Assessment}

As noted in the previous subsection, the same classification format also included a topical component assigning each talk to one of seven predefined categories. Previous studies have shown that the thematic domain of a TED Talk can substantially influence audience engagement, with certain subjects naturally attracting higher levels of interest and interaction. 

In \cite{fischer2024affect}, TED Talks were categorized into seven topical domains (Health, Cosmos, Mind, Environment, Tech, Society, and Entertainment) based on a semantic network analysis of TED-assigned tags. Specifically, the authors constructed a co-occurrence network of 447 tags and applied the Louvain modularity detection algorithm to identify groups of semantically related tags. Each resulting cluster was labeled according to its dominant theme, producing the seven-topic framework widely used in subsequent analyses.

In the present study, we adopted the same seven-category structure but applied it directly to transcript content rather than relying on metadata. Using the same language-model prompt described earlier, the model was instructed to classify each transcript into exactly one of the seven categories based solely on its linguistic and semantic features. Each transcript was evaluated fifteen times, and the most frequently predicted category across runs was assigned as the final topic label.

Unlike tag-based approaches, this method benefits from contextual understanding: instead of relying on isolated keywords, the model interprets each word within the broader context of the entire transcript. This allows for more accurate and semantically coherent classification, ensuring that topic assignment reflects the meaning of the talk as a whole rather than superficial lexical cues.

To assess classification stability, each TED transcript was independently classified by the LLM 15 times. Due to the stochastic nature of large language models, repeated classifications of identical inputs can yield varying outputs. The modal (most frequent) category across the 15 iterations was selected as the final classification for each transcript. 


As shown in Table~\ref{tab:classification_stability}, classification stability was consistently high across all seven categories (range: 89.18\%–94.96\%), with minimal variation confirming robust reliability regardless of content type. These results demonstrate that the LLM classifications were stable and suitable for subsequent analyses.

\begin{table}[H]
\centering
\caption{Stability of topic classification across 15 repeated LLM runs (N = 1,239).}
\label{tab:classification_stability}
\begin{tabular}{lccc}
\hline
\textbf{Topic Category} & \textbf{Mean Agreement (\%)} & \textbf{N} & \textbf{SD} \\
\hline
Cosmos & 91.97 & 49 & 15.45 \\ 
Entertainment & 92.83 & 172 & 13.70 \\ 
Environment & 93.48 & 185 & 13.60 \\ 
Health & 94.96 & 123 & 10.29 \\ 
Mind & 89.18 & 114 & 15.90 \\ 
Society & 92.77 & 397 & 13.79 \\ 
Tech & 90.37 & 199 & 15.33 \\ 
\hline
\textbf{Total} & \textbf{92.35} & \textbf{1,239} & \textbf{14.04} \\
\hline
\end{tabular}
\end{table}

Taken together, these procedures yielded auxiliary variables capturing both the scientific and topical context of each talk. The high level of classification stability indicates that the LLM-based labeling process was statistically reliable and conceptually robust, providing valid and reproducible descriptors for use as control variables in subsequent analyses.

\subsection{Data Cleaning and Exclusion Criteria}
\label{sec:data-cleaning}

Because the present study relies primarily on transcript-based linguistic analysis, the validity of the results depends on spoken language constituting the primary communicative channel of each talk. While most TED Talks follow a lecture-like format, a small subset consists of performances, musical acts, or visually oriented presentations in which speech plays a secondary or minimal role, rendering transcript-based clarity assessment conceptually inappropriate rather than merely low in quality.

\begin{figure}[htbp]
    \centering
    \includegraphics[width=0.75\textwidth]{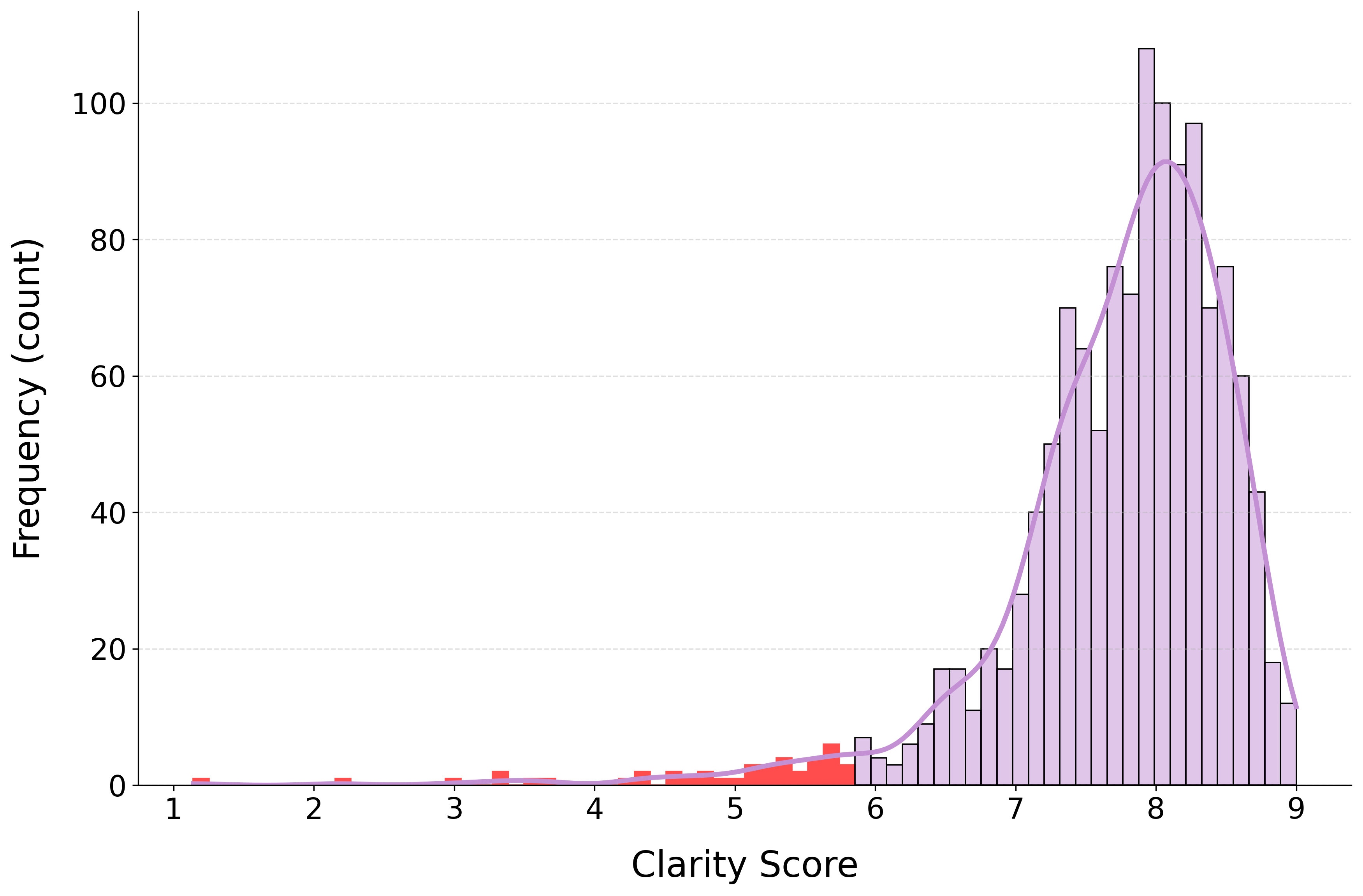}
    \caption{Distribution of AI-derived clarity scores across all TED Talks in the dataset. (N = 1,280).}
    \label{fig:clarity_1280_distribution}
\end{figure}

Inspection of the clarity score distribution (Figure~\ref{fig:clarity_1280_distribution}) reveals a highly concentrated upper range, with the vast majority of talks clustered between approximately 7.0 and 9.0, and a sparse left tail characterized by a sharp drop in frequency density. The lower interquartile-range (IQR) fence of the distribution was 6.08. A conservative cutoff of 5.8 was therefore selected, positioned well within the extreme left tail and below the IQR-based boundary, to ensure that only unequivocal outliers were excluded while avoiding truncation of low-clarity but otherwise valid lectures.

Applying this threshold resulted in the exclusion of 41 talks (3.2\% of the dataset), yielding a final sample of 1,239 talks. Qualitative inspection confirmed that the excluded cases predominantly represented non-lecture formats, including musical performances, visual demonstrations, and fragmented dialogue-based presentations in which linguistic structure was not the dominant mode of communication.

To further assess the validity of this filtering procedure, all 41 excluded talks were independently examined by a blind human reviewer who had no access to the automated clarity scores. Of these, 33 talks (80.5\%) were classified as presentations in which linguistic content did not constitute the dominant communicative modality, including musical performances, visually driven demonstrations, and performative segments.

Importantly, this manual inspection served solely as a post-hoc validation step and did not inform the exclusion decision, which was defined \emph{a priori} based on statistical considerations. This preserves the objectivity and reproducibility of the automated filtering procedure.

To ensure that this exclusion criterion did not artificially influence the results, we conducted all analyses in parallel on both the filtered (N = 1,239) and unfiltered (N = 1,280) datasets. As will be shown in the Results section, the main findings remain substantively unchanged regardless of filtering. Full comparative analyses for the unfiltered dataset are provided in Appendix~\ref{app:full data}.

\subsection{Comparison with Readability-Based Metrics}

To further examine the validity and distinctiveness of the AI-derived clarity measure, we conducted an additional comparison with readability-based metrics reported in prior large-scale analyses of TED Talks \citep{fischer2024affect}. In that study, textual accessibility was quantified using the \textit{Flesch Reading Ease} score \citep{flesch1948new}, a widely used readability metric based primarily on sentence length and word complexity. Higher scores indicate easier-to-read text, whereas lower scores reflect greater linguistic complexity.

To enable a direct comparison between readability and AI-derived clarity, we identified the subset of TED Talks shared across both datasets, resulting in 928 overlapping talks. After applying the same clarity-based exclusion criterion used in the main analysis (see Section~\ref{sec:data-cleaning}), the final comparison sample consisted of 911 TED Talks.

This shared subset enabled a direct comparison between traditional readability metrics and the AI-derived clarity measure, allowing us to assess whether clarity captures communicative qualities beyond those reflected by conventional readability formulas.

\subsection{Data Analysis Tools}

All statistical analyses were conducted using IBM SPSS Statistics and R.  
SPSS was used primarily for descriptive statistics, correlations, and regression models, while R was employed for data preprocessing, visualization, and validation of analytical results.  
Using both platforms allowed for cross-verification of findings and ensured robustness in data handling and statistical inference.

\section{Results}

\subsection{Descriptive Statistics of AI-Derived Scores}

Before examining the relationships among the study variables, descriptive statistics were computed for the AI-generated \textit{clarity} and \textit{structure} scores across all 1,239 TED Talk transcripts.
As shown in Table~\ref{tab:clarity_structure_descriptives}, the AI-derived \textit{clarity} and \textit{structure} scores ranged from approximately 6 to 9 across talks, with mean values of 7.84 and 8.37, respectively.
The corresponding standard deviations were 0.61 for \textit{clarity} and 0.65 for \textit{structure}.
The entire distribution of \textit{clarity} scores reflects our decision to begin the scale at 5.8, which shifts all values upward and limits the lower range.

\begin{table}[H]
\centering
\caption{Descriptive statistics of AI-derived clarity and structure scores (N = 1,239).}
\label{tab:clarity_structure_descriptives}
\begin{tabular}{lccccc}
\hline
\textbf{Variable} & \textbf{N} & \textbf{Minimum} & \textbf{Maximum} & \textbf{Mean} & \textbf{Std. Deviation} \\
\hline
Clarity & 1,239 & 5.82 & 9.00 & 7.84 & 0.61 \\
Structure & 1,239 & 5.18 & 9.36 & 8.37 & 0.65 \\
\hline
\end{tabular}
\end{table}

\subsection{Distribution of Scientific and Topical Classifications}

Each TED Talk was classified by the AI model as either scientific or non-scientific and assigned to one of seven topical categories.  
Out of the 1,239 talks in the dataset, 397 (32\%) were classified as scientific and 842 (68\%) as non-scientific.  
Table~\ref{tab:science_topic_distribution} presents the cross-tabulation of scientificness by topic category.

\begin{table}[H]
\centering
\caption{Distribution of scientific and non-scientific talks across topic categories (N = 1,239).}
\label{tab:science_topic_distribution}
\begin{tabular}{lrrrrrr}
\hline
\multirow{2}{*}{\textbf{Topic}} & \multicolumn{2}{c}{\textbf{Non-scientific}} & \multicolumn{2}{c}{\textbf{Scientific}} & \multicolumn{2}{c}{\textbf{Total}} \\
\cline{2-7}
 & \textbf{N} & \textbf{\%} & \textbf{N} & \textbf{\%} & \textbf{N} & \textbf{\%} \\
\hline
Cosmos & 12 & 1.4 & 37 & 9.3 & 49 & 4.0 \\
Entertainment & 172 & 20.4 & 0 & 0.0 & 172 & 13.9 \\
Environment & 85 & 10.1 & 100 & 25.2 & 185 & 14.9 \\
Health & 36 & 4.3 & 87 & 21.9 & 123 & 9.9 \\
Mind & 55 & 6.5 & 59 & 14.9 & 114 & 9.2 \\
Society & 369 & 43.8 & 28 & 7.1 & 397 & 32.0 \\
Tech & 113 & 13.4 & 86 & 21.7 & 199 & 16.1 \\
\hline
\textbf{Total} & \textbf{842} & \textbf{100.0} & \textbf{397} & \textbf{100.0} & \textbf{1,239} & \textbf{100.0} \\
\hline
\end{tabular}
\end{table}

Overall, scientific talks were most prevalent within the \textit{Environment}, \textit{Health}, and \textit{Tech} domains, whereas non-scientific talks were dominant in the \textit{Society} and \textit{Entertainment} categories.  
Notably, none of the talks classified under \textit{Entertainment} were labeled as scientific.

\subsection{Correlation Matrix}

Table~\ref{tab:correlations} presents Pearson correlation coefficients among the key study variables, including AI-derived \textit{clarity} and \textit{structure} scores, video duration, and engagement metrics (log-transformed views and likes), based on \(N = 1{,}239\) TED Talks.

As shown, \textit{clarity} and \textit{structure} were highly correlated (\(r = .907, p < .01\)), indicating substantial conceptual and linguistic overlap between the two AI-derived dimensions. Because of this strong association, and to avoid potential multicollinearity in subsequent regression analyses, \textit{clarity} was selected as the primary predictor variable for all further statistical comparisons. Conceptually, clarity inherently captures organizational and structural qualities of speech, meaning that variation in structure is largely embedded within the broader construct of clarity.

A clear pattern also emerges in which \textit{clarity} shows one of the strongest associations with audience engagement metrics. Specifically, \textit{clarity} was positively correlated with both \textit{Likes} (\(r = .373, p < .01\)) and \textit{Views} (\(r = .316, p < .01\)), indicating that talks rated as clearer tended to receive more likes and views. \textit{Structure} also showed positive but slightly weaker correlations with these engagement indicators (\(r = .290\) with \textit{Likes}; \(r = .228\) with \textit{Views}). In contrast, \textit{Duration} was weakly and positively correlated with engagement, suggesting that longer talks attracted somewhat more views and likes, though the effect size was minimal. Finally, the correlation between \textit{Views} and \textit{Likes} was extremely high (\(r = .964\)), reflecting their near-overlapping nature as indicators of audience response.

A parallel non-parametric analysis using Spearman rank-order correlations (see Appendix~\ref{app:spearman}) yielded a highly similar pattern of associations. This convergence supports the assumption that the relationships among the variables are sufficiently monotonic and approximately linear, thereby justifying the use of Pearson correlations and linear regression models in the main analyses.

\begin{table}[H]
\centering
\caption{Pearson correlation coefficients among key variables (N = 1,239).}
\label{tab:correlations}
\begin{adjustbox}{max width=\textwidth}
\begin{tabular}{lcccccc}
\toprule
\textbf{Variable} & TED\_TrendIndex & Clarity & Structure & Duration (s) & Views & Likes \\
\midrule
TED\_TrendIndex & 1 & .278** & .272** & -.255** & .173** & .280** \\
Clarity &  & 1 & .907** & -.089** & .316** & .373** \\
Structure &  &  & 1 & -.070* & .228** & .290** \\
Duration (s) &  &  &  & 1 & .109** & .104** \\
Views &  &  &  &  & 1 & .964** \\
Likes &  &  &  &  &  & 1 \\
\bottomrule
\end{tabular}
\end{adjustbox}
\begin{flushleft}
\textit{Note.} *\(p < .05\), **\(p < .01\) (two-tailed).
\end{flushleft}
\end{table}

\subsection{Hierarchical Regression Predicting Likes}
\label{Reg1-3}

To assess the predictive role of AI-derived linguistic clarity, a three-step hierarchical regression was conducted with \textit{Likes} as the dependent variable (Table~\ref{tab:reg_likes}).

In Step~1, two baseline predictors were entered: the the TED Google Trends index at the time of release and talk duration. Together, they explained 11.1\% of the variance ($R^2 = .111$, $F = 77.53$, $p < .001$). Both predictors were significant, with \textit{TED\_TrendIndex} showing a relatively strong effect ($\beta = .328$, $p < .001$) and \textit{Duration} also contributing positively ($\beta = .188$, $p < .001$). This indicates that contextual visibility and basic exposure-related factors account for a meaningful initial portion of liking behavior, but leave most of the variance unexplained.

In Step~2, the scientific classification and topic categories were added, increasing explained variance to 19.5\% ($\Delta R^2 = .083$, $F = 32.98$, $p < .001$). Several topic effects were substantial: talks in the \textit{Mind} category received more likes relative to \textit{Society} ($\beta = .202$), whereas \textit{Health} ($\beta = -.109$) and \textit{Environment} ($\beta = -.132$) received fewer likes. \textit{Entertainment} talks were also positively associated with likes ($\beta = .052$). The scientific indicator showed only a small effect ($\beta = .064$, $p < .05$), suggesting a weak and potentially unstable advantage at this stage.

In Step~3, the AI-derived \textit{Clarity} score was introduced, leading to a marked improvement in explanatory power. The model explained 29\% of the variance ($\Delta R^2 = .095$, $F = 50.10$, $p < .001$). \textit{Clarity} emerged as the strongest predictor in the full model ($\beta = .339$, $p < .001$), indicating that clearer transcripts are strongly associated with higher levels of audience appreciation. Importantly, once clarity was included, the scientific classification became entirely non-significant ($\beta = .018$, $p = .55$), suggesting that the small initial scientific advantage was fully accounted for by linguistic factors. Topic effects such as the positive coefficient for \textit{Mind} ($\beta = .189$) and the negative effects for \textit{Health} ($\beta = -.089$) and \textit{Environment} ($\beta = -.108$) remained robust.

Overall, clarity explained the largest share of incremental variance beyond contextual and thematic predictors. This supports the view that linguistic clarity represents a central communicative cue shaping audience engagement, over and above both exposure conditions and content category.

\begin{table}[htbp]
\centering
\caption{Hierarchical regression predicting \textit{Likes} from temporal, thematic, and linguistic predictors.}
\label{tab:reg_likes}
\begin{tabular}{lllllll}
\toprule
 & \textbf{Predictor} & \textbf{$\beta$} & \textbf{t} & \textbf{F} & \textbf{$R^2$} & \textbf{$\Delta R^2$} \\
\midrule

\textbf{Step I} & TED\_TrendIndex & 0.328 & 11.83$^{***}$ & 77.53$^{***}$ & 0.111  \\
 & Duration (s) & 0.188 & 6.77$^{***}$ \\
\midrule

\textbf{Step II} & TED\_TrendIndex & 0.315 & 11.75$^{***}$ & 32.98$^{***}$ & 0.195 & 0.083 \\
 & Duration (s) & 0.165 & 6.11$^{***}$ \\
 & Science & 0.064 & 2.05$^{*}$ \\
 & Cosmos (vs. Society) & -0.014 & -0.48 \\
 & Mind (vs. Society) & 0.202 & 6.96$^{***}$ \\
 & Tech (vs. Society) & -0.072 & -2.37$^{*}$ \\
 & Entertainment (vs. Society) & 0.052 & 1.82 \\
 & Health (vs. Society) & -0.109 & -3.56$^{***}$ \\
 & Environment (vs. Society) & -0.132 & -4.29$^{***}$ \\
\midrule

\textbf{Step III} & TED\_TrendIndex & 0.229 & 8.79$^{***}$ & 50.10$^{***}$ & 0.290 & 0.095 \\
 & Duration (s) & 0.187 & 7.36$^{***}$ \\
 & Science & 0.018 & 0.60 \\
 & Cosmos (vs. Society) & -0.006 & -0.21 \\
 & Mind (vs. Society) & 0.189 & 6.92$^{***}$ \\
 & Tech (vs. Society) & -0.026 & -0.90 \\
 & Entertainment (vs. Society) & 0.139 & 5.02$^{***}$ \\
 & Health (vs. Society) & -0.089 & -3.09$^{**}$ \\
 & Environment (vs. Society) & -0.108 & -3.73$^{***}$ \\
 & Clarity & 0.339 & 12.83$^{***}$ \\
\bottomrule

\multicolumn{7}{l}{\small Note: $^{*}p<.05$; $^{**}p<.01$; $^{***}p<.001$.} \\
\end{tabular}
\end{table}

\subsection{Hierarchical Regression Predicting Views}

To examine whether a similar pattern holds for content exposure, a parallel three-step hierarchical regression was conducted with \textit{Views} as the dependent variable (Table~\ref{tab:reg_views}).

In Step~1, the TED Google Trends index and talk duration jointly explained 5.5\% of the variance in views ($R^2 = .055$, $F = 36.12$, $p < .001$). Both predictors were significant: higher real-time public interest was associated with more views ($\beta = .215$, $p < .001$), and longer talks also accumulated more views ($\beta = .164$, $p < .001$). Compared to the Likes model, baseline predictors accounted for a smaller portion of variance, suggesting that viewing behavior is influenced by additional factors not captured by these two variables.

In Step~2, the scientific flag and topic categories were added, increasing explained variance to 14.3\% ($\Delta R^2 = .088$, $F = 22.86$, $p < .001$). Strong topic effects emerged: talks in the \textit{Mind} category received more views relative to \textit{Society} ($\beta = .212$), while \textit{Health} ($\beta = -.103$) and \textit{Environment} ($\beta = -.136$) received fewer views. The scientific indicator did not reach significance ($\beta = .051$, $p = .116$), indicating that scientific labeling alone does not meaningfully affect exposure levels.

In Step~3, adding the AI-derived \textit{clarity} score further improved the model to 22.5\% explained variance ($\Delta R^2 = .082$, $F = 35.67$, $p < .001$). Clarity emerged as a strong positive predictor of views ($\beta = .314$, $p < .001$), demonstrating that talks with clearer linguistic structure tend to reach larger audiences even after controlling for timing, duration, topic, and scientific classification. The scientific flag remained non-significant ($\beta = .008$), and topic effects remained stable.

Taken together, clarity contributes substantial incremental explanatory power for views as well, although its standardized effect is slightly smaller than for likes. This suggests that while views are more strongly shaped by external platform dynamics and exposure conditions, linguistic clarity still plays a central and independent role in determining how widely a talk is consumed.

\begin{table}[htbp]
\centering
\caption{Hierarchical regression predicting \textit{Views} from temporal, thematic, and linguistic predictors.}
\label{tab:reg_views}
\begin{tabular}{lllllll}
\toprule
\textbf{Step} & \textbf{Predictor} & \boldmath{$\beta$} & \textbf{t} & \textbf{F} & \boldmath{$R^2$} & \boldmath{$\Delta R^2$} \\
\midrule

\textbf{I} 
& TED\_TrendIndex & 0.215 & 7.52$^{***}$ & 36.12$^{***}$ & 0.055 &  \\
& Duration (s) & 0.164 & 5.74$^{***}$ &  &  &  \\
\midrule

\textbf{II} 
& TED\_search & 0.202 & 7.28$^{***}$ & 22.86$^{***}$ & 0.143 & 0.088 \\
& Duration (s) & 0.145 & 5.23$^{***}$ &  &  &  \\
& Science & 0.051 & 1.57 &  &  &  \\
& Cosmos (vs. Society) & -0.018 & -0.61 &  &  &  \\
& Mind (vs. Society) & 0.212 & 7.08$^{***}$ &  &  &  \\
& Tech (vs. Society) & -0.052 & -1.67 &  &  &  \\
& Entertainment (vs. Society) & 0.070 & 2.36$^{*}$ &  &  &  \\
& Health (vs. Society) & -0.103 & -3.25$^{***}$ &  &  &  \\
& Environment (vs. Society) & -0.136 & -4.29$^{***}$ &  &  &  \\
\midrule

\textbf{III} 
& TED\_search & 0.122 & 4.48$^{***}$ & 35.67$^{***}$ & 0.225 & 0.082 \\
& Duration (s) & 0.166 & 6.25$^{***}$ &  &  &  \\
& Science & 0.008 & 0.25 &  &  &  \\
& Cosmos (vs. Society) & -0.010& -0.38 &  &  &  \\
& Mind (vs. Society) & 0.199 & 7.01$^{***}$ &  &  &  \\
& Tech (vs. Society) & -0.010 & -0.33 &  &  &  \\
& Entertainment (vs. Society) & 0.150 & 5.19$^{***}$ &  &  &  \\
& Health (vs. Society) & -0.084 & -2.80$^{**}$ &  &  &  \\
& Environment (vs. Society) & -0.114 & -3.77$^{***}$ &  &  &  \\
& Clarity & 0.314 & 11.38$^{***}$ &  &  &  \\
\bottomrule

\multicolumn{7}{l}{\small Note: $^{*}p<.05$; $^{**}p<.01$; $^{***}p<.001$.} \\
\end{tabular}
\end{table}

\subsection{Interaction Effects}

Because scientific content classification did not significantly predict engagement in Model~2, we examined whether the predictive effect of clarity varies across topical domains or between scientific and non-scientific talks. Although categories showed some variation in their average clarity levels ($M = 7.45$–$8.09$) and engagement ($M = 3.13$–$3.94$), the key question was whether clarity functions differently across domains.

To assess this, we estimated a fourth hierarchical regression model (Model~4) adding all Category~$\times$~Clarity interactions as well as the Science~$\times$~Clarity interaction. Most terms did not reach statistical significance, and the interaction block explained negligible incremental variance ($\Delta R^{2} = .006$).

To further validate this invariance, clarity–engagement correlations were computed separately for each topic category. The correlations were positive across all domains and statistically significant in most categories, with comparable effect sizes across topics (see Appendix~\ref{app:category_descriptives}). These results indicate that the effect of clarity is broadly consistent across TED content types.

A full presentation of Model~4 appears in Appendix~\ref{app:model4}.

\subsection{Validation Across Prompt Versions and Models}
\label{subsec:Validation}

To assess the robustness and generalizability of the AI-based evaluation approach, we conducted a validation analysis comparing several configurations: the TED-focused prompt using GPT-4o (the primary evaluation method used throughout this study), an academic lecture prompt using GPT-4o, and the TED-focused prompt applied to alternative state-of-the-art language models (Claude 3.5 Sonnet and Gemini 3 Flash; for full prompt text, see Appendix~\ref{app:prompt}). This comparison served two purposes: first, to examine whether context-specific prompt design yields meaningfully different results across educational settings (TED talks versus university lectures); and second, to evaluate cross-model consistency using the same TED-focused prompt.

For this validation analysis, we selected all TED Talks published in 2010 as a representative subsample. The year 2010 was chosen because it falls near the midpoint of the dataset's temporal range (2006–2013), ensuring adequate representation of TED's established presence on YouTube while providing a sufficiently large sample for reliable statistical comparisons.

Because language models occasionally declined to evaluate certain talks (e.g., non-verbal performances or content restricted by safety policies), the effective sample size varied slightly across models (approximately N = 184–190). To ensure comparability and avoid bias, all correlations were computed using the overlapping subset of talks for which valid scores were available for each model.

Table~\ref{tab:prompt_model_comparison} presents the Pearson correlation coefficients among engagement metrics and the \textit{Clarity} and \textit{Structure} scores derived from each evaluation configuration.

The results reveal several key findings. First, the TED-focused prompt using GPT-4o produced substantially stronger associations with engagement metrics than the academic lecture prompt. Specifically, \textit{Clarity} scores from the TED prompt correlated with \textit{Likes} at $r = .390$ and with \textit{Views} at $r = .386$, whereas the academic prompt yielded weaker correlations of $r = .219$ with \textit{Likes} and $r = .199$ with \textit{Views}.

Second, the structure dimension from the academic lecture prompt showed weak and non-significant associations with engagement ($r = .131$ with \textit{Likes}; $r = .110$ with \textit{Views}), whereas structure scores from the TED prompt remained significant, albeit weaker than \textit{Clarity} ($r = .262$ with \textit{Likes}; $r = .261$ with \textit{Views}).

Third, alternative models using the TED-focused prompt produced correlations that were generally intermediate but consistently positive. Claude 3.5 Sonnet yielded \textit{Clarity} correlations of $r = .248$ with \textit{Likes} and $r = .219$ with \textit{Views}, and \textit{Structure} correlations of $r = .236$ and $r = .234$, respectively. Gemini 3 Flash produced \textit{Clarity} correlations of $r = .271$ with \textit{Likes} and $r = .270$ with \textit{Views}, while \textit{Structure} correlations were slightly higher ($r = .295$ with \textit{Likes} and $r = .303$ with \textit{Views}).

Fourth, across most models and prompt configurations, \textit{Clarity} outperformed \textit{Structure} in predicting engagement. In addition, the \textit{Clarity} and \textit{Structure} scores produced by different models were strongly intercorrelated, indicating substantial convergence in their assessment of communicative quality despite architectural and policy differences.

Taken together, these findings demonstrate that genre-aligned prompt design substantially improves predictive validity, and that the relationship between transcript-based clarity and audience engagement is robust across multiple state-of-the-art language models.

\begin{table}[htbp]
\centering
\caption{Pearson correlations among engagement metrics and AI-derived scores across prompt types and models (year 2010).}
\label{tab:prompt_model_comparison}

\resizebox{\textwidth}{!}{%
\begin{tabular}{lcccccccccc}
\hline
\textbf{Variable} & \textbf{Likes} & \textbf{Views} & \textbf{C\_GPT4o} & \textbf{S\_GPT4o} & \textbf{C\_GPT4o} & \textbf{S\_GPT4o} & \textbf{C\_Claude} & \textbf{S\_Claude} & \textbf{C\_Gemini} & \textbf{S\_Gemini} \\
 &  &  & \textbf{TED} & \textbf{TED} & \textbf{Academic} & \textbf{Academic} & \textbf{TED} & \textbf{TED} & \textbf{TED} & \textbf{TED} \\
\hline
Likes & 1 & .969** & .390** & .262** & .219** & .131 & .248** & .236** & .271** & .295** \\
Views &  & 1 & .386** & .261** & .199** & .110 & .219** & .234** & .270** & .303** \\
C\_GPT4o\_TED &  &  & 1 & .904** & .817** & .745** & .666** & .571** & .659** & .656** \\
S\_GPT4o\_TED &  &  &  & 1 & .814** & .802** & .628** & .561** & .673** & .652** \\
C\_GPT4o\_Academic &  &  &  &  & 1 & .958** & .622** & .562** & .590** & .608** \\
S\_GPT4o\_Academic &  &  &  &  &  & 1 & .557** & .532** & .530** & .583** \\
C\_Claude\_TED &  &  &  &  &  &  & 1 & .699** & .592** & .537** \\
S\_Claude\_TED &  &  &  &  &  &  &  & 1 & .523** & .490** \\
C\_Gemini\_TED &  &  &  &  &  &  &  &  & 1 & .776** \\
S\_Gemini\_TED &  &  &  &  &  &  &  &  &  & 1 \\
\hline
\end{tabular}%
}

\begin{flushleft}
\textit{Note.} $^{**}p<.01$ (two-tailed).  
C = Clarity; S = Structure.  
Effective sample sizes varied slightly across models (approximately $N = 184$–$190$) due to occasional model refusals; correlations were computed using the available overlapping data.
\end{flushleft}

\end{table}

\subsection{Temporal Trends in Clarity and Phase-Based Segmentation}

\begin{figure}[t]
    \centering
    \includegraphics[width=0.75\linewidth]{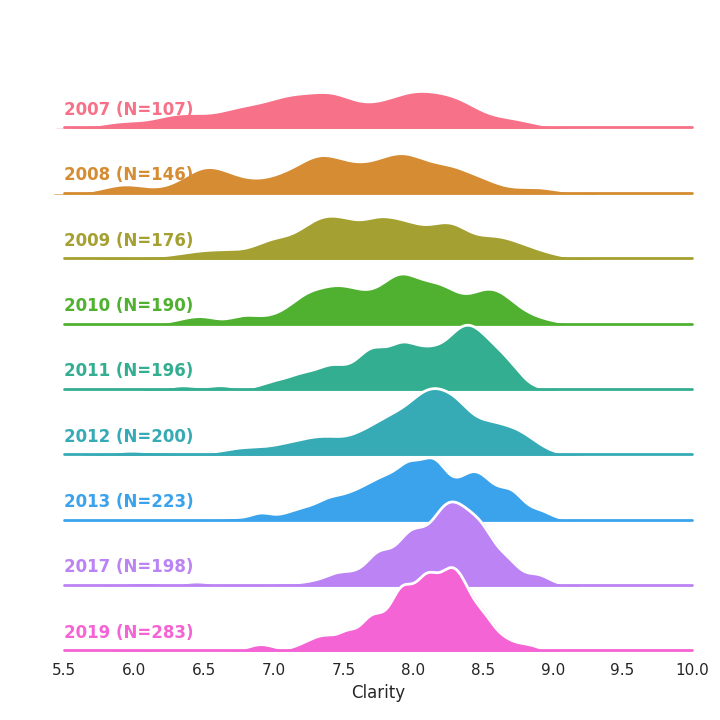}
    \caption{Ridgeline density plots of clarity scores by year. Each curve represents the distribution of clarity values for a given year, with sample size indicated in parentheses. Over time, the distributions shift rightward and become increasingly concentrated, reflecting a steady increase in mean clarity and a concurrent decrease in variability.}
    \label{fig:clarity_temporal_distributions}
\end{figure}

Figure~\ref{fig:clarity_temporal_distributions} and Table~\ref{tab:clarity_structure_temporal_descriptive} jointly illustrate a pronounced temporal trend in clarity scores across TED Talks. Over the examined period, the mean clarity score exhibits a consistent increase, accompanied by a systematic reduction in standard deviation. This pattern indicates not only an overall improvement in communicative clarity but also a progressive homogenization of presentation quality across talks.

As illustrated in Figure~\ref{fig:clarity_temporal_distributions}, early years are characterized by broad and heterogeneous distributions, spanning a wide range of clarity values. In contrast, later years exhibit markedly narrower distributions, concentrated around high clarity scores. This rightward shift and progressive narrowing indicate a transition from substantial inter-talk variability toward a regime of high communicative consistency.

These trends are quantitatively summarized in Table~\ref{tab:clarity_structure_temporal_descriptive}, which reports the yearly mean, standard deviation, and sample size of \textit{Clarity} and \textit{Structure}. The table confirms a monotonic increase in average clarity over time, alongside a consistent reduction in dispersion. Together, the visual and numerical analyses suggest that contemporary TED Talks increasingly conform to shared standards of rhetorical structure, linguistic simplicity, and audience-oriented delivery.

\begin{table}[t]
\centering
\caption{Descriptive statistics of clarity and structure scores by year.}
\label{tab:clarity_structure_temporal_descriptive}
\begin{tabular}{lccccc}
\hline
\textbf{Year} & \textbf{Mean Clarity} & \textbf{SD Clarity} & \textbf{Mean Structure} & \textbf{SD Structure} & \textbf{N} \\
\hline
2007 & 7.49 & 0.70 & 7.96 & 0.87 & 107 \\
2008 & 7.47 & 0.71 & 7.95 & 0.85 & 146 \\
2009 & 7.75 & 0.60 & 8.28 & 0.62 & 176 \\
2010 & 7.85 & 0.56 & 8.41 & 0.54 & 190 \\
2011 & 8.01 & 0.49 & 8.53 & 0.51 & 196 \\
2012 & 7.99 & 0.56 & 8.52 & 0.57 & 200 \\
2013 & 8.05 & 0.46 & 8.60 & 0.38 & 223 \\
2017 & 8.15 & 0.44 & 8.71 & 0.41 & 198 \\
2019 & 8.05 & 0.39 & 8.67 & 0.33 & 283 \\
\hline
\end{tabular}
\end{table}

Beyond documenting a positive longitudinal trend, this convergence toward high clarity introduces a new methodological challenge. As clarity scores become increasingly compressed within a narrow high-performance range, discriminating between talks of ``good'' and ``excellent'' clarity becomes substantially more difficult. The reduced variance limits the sensitivity of conventional statistical analyses and necessitates more fine-grained modeling strategies.

Accordingly, we conceptualize the dataset as comprising two distinct phases. The early phase is characterized by high heterogeneity, broad distributions, and large inter-talk variability. The late phase, by contrast, exhibits high homogeneity, elevated mean clarity, and substantially reduced dispersion. The following subsection focuses on the analytical challenges posed by this late phase and outlines methodological adaptations required to maintain discriminative power under conditions of reduced variability.

\subsection{Correlation Structure in the Late Phase}

The late phase of the dataset, spanning the years 2017 and 2019, is characterized by uniformly high clarity scores and substantially reduced variability. This convergence provides an informative setting for examining how linguistic quality metrics relate to audience engagement under conditions of restricted variance.

To ensure methodological consistency with the primary analysis, the low-clarity subset within this period was defined using an objective, distribution-based criterion. Specifically, an interquartile range (IQR) threshold was applied to the clarity scores obtained from the GPT-based model, yielding a cutoff value of 7.21. This threshold corresponds to approximately the lowest 3\% of the clarity distribution, closely matching the exclusion proportion employed in the main analysis.

Because models occasionally declined to evaluate specific talks (e.g., non-verbal performances or policy-restricted content), effective sample sizes varied slightly across models (approximately $N = 458$–$468$). To ensure comparability, correlations were computed using the overlapping subset of talks with valid evaluations.

Table~\ref{tab:late_phase_corr} presents the Pearson correlation coefficients between engagement metrics and AI-derived scores across configurations.

As expected, \textit{Views} and \textit{Likes} exhibited an almost perfect correlation ($r = .972$, $p < .01$), reflecting their shared role as closely related indicators of audience engagement.

More importantly, the associations between linguistic quality measures and engagement were substantially attenuated relative to the primary dataset. GPT-derived \textit{Clarity} showed modest but statistically significant correlations with both \textit{Likes} ($r = .144$, $p < .01$) and \textit{Views} ($r = .136$, $p < .01$), whereas GPT-based \textit{Structure} was not significantly related to either engagement metric ($r = .081$ for both outcomes). 

In contrast, the Gemini-based evaluations produced somewhat stronger associations. \textit{Clarity} from Gemini correlated with \textit{Likes} and \textit{Views} at $r = .254$ ($p < .01$), while \textit{Structure} showed the strongest relationships in this phase ($r = .343$ with \textit{Likes}$;$ $r = .336$ with \textit{Views}$,$ both $p < .01$). Despite these differences in magnitude, all significant effects remained positive, indicating that higher linguistic quality continued to be associated with greater audience engagement.

Correlations among linguistic variables remained high across models, suggesting strong cross-model convergence even under restricted variance conditions. GPT-derived \textit{Clarity} and \textit{Structure} were strongly related ($r = .850$, $p < .01$), and substantial correlations were also observed between GPT and Gemini measures (e.g., $r = .436$ between clarity scores).

Taken together, these results indicate that the clarity–engagement relationship persists in the late phase but is markedly weaker than in the earlier, more heterogeneous period. This attenuation is consistent with a statistical range-restriction effect: when clarity scores cluster within a narrow high-performance band, the ability to discriminate between talks diminishes, reducing observable correlations with behavioral outcomes. In other words, once communicative quality becomes uniformly high, additional gains in clarity yield progressively smaller differences in audience response.

Importantly, the persistence of positive correlations across models suggests that linguistic quality continues to function as a meaningful predictor of engagement even under conditions approaching a ceiling. The late-phase results therefore support the interpretation that the declining effect size reflects reduced variability rather than a substantive weakening of the underlying relationship between clarity and audience engagement.

\begin{table}[t]
\centering
\caption{Pearson correlation matrix for linguistic quality measures and audience engagement in the late phase (2017 and 2019).}
\label{tab:late_phase_corr}
\resizebox{\linewidth}{!}{%
\begin{tabular}{lcccccc}
\hline
 & \textbf{Likes} & \textbf{Views} & \textbf{Clarity GPT} & \textbf{Structure GPT} & \textbf{Clarity Gemini} & \textbf{Structure Gemini} \\
\hline
Likes        & 1 & .972$^{**}$ & .144$^{**}$ & .081 & .254$^{**}$ & .343$^{**}$ \\
Views        &        & 1 & .136$^{**}$ & .081 & .254$^{**}$ & .336$^{**}$ \\
Clarity GPT       &        &        & 1 & .850$^{**}$ & .436$^{**}$ & .432$^{**}$ \\
Structure GPT     &        &        &        & 1 & .483$^{**}$ & .442$^{**}$ \\
Clarity Gemini    &        &        &        &        & 1 & .681$^{**}$ \\
Structure Gemini  &        &        &        &        &        & 1 \\
\hline
\end{tabular}
}
\begin{flushleft}
\textit{Note.} $^{**}p < .01$ (two-tailed).
\end{flushleft}
\end{table}

\subsection{Comparison with Readability-Based Metrics}

To further examine whether AI-derived clarity captures communicative qualities beyond traditional readability measures, we computed Pearson correlations between \textit{Clarity}, the \textit{Flesch Reading Ease} readability score reported by \cite{fischer2024affect}, and audience engagement metrics (\textit{Views} and \textit{Likes}) using the shared dataset ($N = 911$).

As shown in Table~\ref{tab:clarity_readability_correlations}, AI-derived \textit{Clarity} demonstrated substantially stronger associations with audience engagement than the readability metric. Specifically, \textit{Clarity} correlated with \textit{Likes} at $r = .364$ ($p < .01$) and with \textit{Views} at $r = .324$ ($p < .01$). In contrast, the \textit{Flesch Reading Ease} readability score showed weaker correlations with \textit{Likes} ($r = .162$, $p < .01$) and \textit{Views} ($r = .187$, $p < .01$).

Interestingly, \textit{Clarity} and \textit{Readability} were weakly negatively correlated ($r = -.120$, $p < .01$), suggesting that AI-derived clarity captures communicative qualities that are largely distinct from traditional readability metrics. While readability primarily reflects surface-level linguistic properties such as sentence length and word complexity, the AI-derived clarity measure appears to capture higher-level discourse characteristics, including explanatory coherence and logical organization.

Taken together, these results indicate that AI-derived clarity provides a stronger and conceptually richer predictor of audience engagement than conventional readability formulas, supporting the added value of AI-based discourse evaluation.

\begin{table}[htbp]
\centering
\caption{Pearson correlations between AI-derived clarity, readability (Flesch Reading Ease), and engagement metrics (N = 911).}
\label{tab:clarity_readability_correlations}
\begin{tabular}{lcccc}
\hline
\textbf{Variable} & \textbf{Clarity} & \textbf{Readability} & \textbf{Views} & \textbf{Likes} \\
\hline
Clarity & 1 & -.120** & .324** & .364** \\
Readability &  & 1 & .187** & .162** \\
Views &  &  & 1 & .968** \\
Likes &  &  &  & 1 \\
\hline
\multicolumn{5}{l}{\small Note: **$p < .01$ (two-tailed).}
\end{tabular}
\end{table}

\section{Discussion}

The current study utilized Large Language Models (LLMs) to investigate the linguistic drivers of digital engagement in science communication. By analyzing over 1,200 TED Talk transcripts, we found that linguistic clarity is a robust predictor of audience engagement. Our findings indicate that high-clarity talks consistently garner more likes and views, suggesting that clarity is a unique and powerful driver in the attention economy.

\subsection{Discussion of Main Findings}

Our primary finding is the dominant role of clarity in predicting engagement metrics. In hierarchical regression models, clarity was the strongest predictor for both likes and views, explaining significant variance above and beyond talk duration, topic, and across both time-frames ($\beta \approx .34$). These results support our main hypothesis, which was that the way information is structured and explained impacts audience engagement. In fact, clarity turned out to have as much impact on the audience as the subject matter itself. Given that the subject matter is usually fixed for most experts, speakers are well-advised to focus on what can be controlled, namely the communicative "packaging" or "engineering" of the message to optimize clarity \citep{sugimoto2013scientists, doumont2009trees}.

Another noteworthy result is that the effect of clarity held up across different academic and non-academic domains. Whether speakers were discussing high-level physics or personal development, the audience's propensity to "like" the content was predicted by the linguistic accessibility of the text transcript (but see discussion below regarding how the context, such as a physics TED talk vs. a physics university lecture, would matter). This suggests a universal preference for clear content in digital environments, where cognitive resources are often limited and attention is fleeting.

\subsection{Comparison with Readability-Based Approaches}
A relevant point of comparison emerges when considering prior readability-based analyses of TED Talks, particularly the work of \citet{fischer2024affect}, which relied on traditional readability metrics such as the Flesch Reading Ease score. These measures primarily capture surface-level linguistic properties, including sentence length and word complexity, and have been widely used as proxies for text accessibility.
The present findings extend this line of research by demonstrating that AI-derived clarity captures a deeper level of communicative quality, and that it showed only weak correlation with traditional readability metrics. Whereas readability metrics focus on lexical and syntactic simplicity, the AI-based clarity measure reflects higher-order discourse features, including explanatory coherence, logical organization, and conceptual flow. Consistent with this distinction, AI-derived clarity demonstrated substantially stronger associations with audience engagement than traditional readability metrics. Taken together, by capturing holistic communicative structure rather than surface linguistic features, AI-based clarity measures provide a richer and more predictive framework for understanding audience engagement in digital communication environments.

\subsection{Theoretical Implications}
Theoretically, these results align well with findings from decades of research on public speaking, teacher education, and science communication. In particular, our results support the Processing Fluency framework \citep{bullock2021narratives, bullock2019jargon, reber2004processing} and could be seen as the "spoken-language" equivalent of the so-called "simple-writing heuristic" identified for large-scale online texts \citep{shulman2024reading}. In brief, processing fluency posits that information that is easy to process is not only better understood but also more positively evaluated. 

More broadly, this aligns with canonical results from journalism, mass communication, and entertainment research \citep{schramm1957twenty, flesch1948new, csikszentmihalyi91flow}. For example, in the context of newspaper reading, classical work by Schramm’s (1954) on the "fraction of selection" suggests that the probability of a person selecting a message is determined by the expectation of reward divided by the effort required. By increasing clarity, speakers decrease the "effort" denominator, thereby increasing the likelihood of selection or sustained attention (i.e. not stopping but continuing to read an article); relatedly, the notion of flow suggests that when task difficulty and person ability are matched, optimal engagement results \citep{csikszentmihalyi91flow}; and lastly, recent work on computational modeling of media choices shows that various content (like genre) and person factors (like mood) can forecast individual choice \citep{gonghuskey2024modeling}, such as whether to switch or keep consuming content. Our work speaks to this by showing that clarity of content is a key driver of collective audience choices and preferences. 

The picture that emerges from our study is one in which clarity promotes fluency or ease of processing, which are metacognitive experiences. These experiences involve more positive/less negative implicit affect, which gets consulted heuristically to inform decisions about engagement (i.e. viewing, liking). We note, however, that we did not study these experiences experimentally and at an individual level (where the impacts are also likely too weak to be measureable), but at the level of aggregate audiences and collective mass decisions. However, the work on processing fluency mentioned above \citep{shulman2024reading} as well as previous work on jargon \citep{oppenheimer2006consequences} support this reasoning. Specifically, Oppenheimer demonstrated that authors who use unnecessarily complex language are often judged as less intelligent and their work as lower quality. Our data show the negative consequences of technical density and linguistic complexity may not only impact person evaluations, but even scale to the public sphere, serving as a barrier to engagement. Thus, extending previous lab-work to the public sphere, the positive correlation between clarity and engagement supports the idea that clear explanations reduce the cognitive load \citep{sweller1988cognitive} on the viewer. Given that high cognitive load tends to be experienced as aversive \citep{david2024unpleasant}, the reduced mental effort due to higher clarity should - all else being equal - make a talk more pleasant. It is important to note that by and large the TED talks were already high in clarity and became even clearer with time. This leads us to expect that in more mundane, less trained and prepared talk settings (e.g. classroom education, public speaking), the effects of clarity are likely even stronger. 

\subsection{Practical Implications: Optimizing Communication, Augmenting Speakers, and Improving Education}

Our findings have clear and actionable implications for science communicators, public speakers, and educators: Clarity must not be treated as a secondary byproduct of expertise, but as the primary goal of the presentation. Our results suggest that speakers can significantly increase their engagement potential by prioritizing structural organization and explanatory quality. This is also supported by a collective body or research across educational psychology, communication science and public speaking training, or even classical signal/noise processing perspectives in engineering. Starting with the latter, from a signal processing perspective, clarity optimizes the signal-to-noise ratio, ensuring the intended message reaches the receiver with minimal distortion \citep{shannon1948mathematical}. 

This technical optimization perspective is also mirrored in the timeless writing advice of Strunk and White \citep{strunkwhite1999style}, whose mandate to "omit needless words" and "make every word tell" remains among most effective strategies for managing audience attention; and in Grice's famous maxims for conversational pragmatics, clarity also features prominently under the label "manner" \citep{grice1975maxims}. In educational psychology, this aligns with the need for high local and global cohesion to facilitate reading comprehension \citep{mcnamara2013reading, mcnamara1996formal}. In the public speaking literature, there seems to be less emphasis on clarity (but see \citep{doumont2009trees} for a strong counterpoint) - even though the concept is touched upon by the notion of "logos" in classical rhetoric or "source expertise/source credibility" \citep{cicero1942deoratore, aristotle2013poetics}. In sum, across a diverse body of work in psychology, communication, and education, clarity emerges as a key factor, leading to the follow-up question of how clarity could be optimized?

Unfortunately, it seems not to be the case that clarity simply follows from expertise, although a certain level of expertise may be necessary. However, the so-called curse of knowledge \citep{wieman2007curse} and related perspective gaps between the sender (speaker/teacher) and the receiver (student/audience) create a key obstacle. With this in mind, our study not only highlights the utility of LLMs as "clarity-yardsticks" but also suggests them as potentially helpful feedback tools. Specifically, by creating an "algorithmic feedback loop" to bridge the gap between technical expertise and public accessibility, LLMs could also be prompted to improve a talk's clarity. This is similar to the use of e.g. virtual reality for training public speaking skills in the nonverbal and performative domain \citep{kroczek2023speaking, saufnay2024improvingps}, but much simpler and more scalable: Just as writers use spell-checkers, scientists and educators might now use AI to assess the communicative quality of their scripts before recording or publishing. In this vein, direct feedback on the linguistic clarity of the text, presumably one of the easiest and most early-stage preparatory activities, could go a long way to optimize communication.

\subsection{Methodological Considerations: The best "clarity-prompt" and TED talks as a unique genre}
An important methodological implication of the present findings concerns the alignment between evaluation criteria and communicative genre. In this study, we explicitly examined whether transcript-based clarity should be assessed using an academic-lecture framework or a TED-specific communicative framework. Although both prompts targeted similar surface dimensions (clarity and structure), their predictive validity differed: When the academic lecture prompt (originally developed for university-level instruction, \citep{zion2025ai}) was applied to TED Talk transcripts, its correlations with engagement metrics were weaker. In contrast, the TED-adapted prompt yielded stronger and more consistent associations with both likes and views.

Although more research is needed, we do not believe that this pattern is a technical artifact of prompt wording. Rather, we interpret it as evidence that communicative quality is not a context-free attribute and that what constitutes "clarity" depends on the rhetorical norms, audience expectations, and communicative goals of the genre under study. In other words, TED Talks are not academic lectures, they are performance-oriented, highly rehearsed presentations designed for broad and heterogeneous audiences, emphasizing narrative flow, accessibility, and cognitive ease rather than disciplinary density or formal exposition \citep{anderson2016ted}. This aligns again with classical and more modern work on types of public speeches and occasions, such as the distinction between informative, persuasive, and celebratory talks \citep{lucas2020ps}, which goes back to Aristotle's and Cicero's typologies (docere/teach, delectare/entertain, movere/persuade, \citep{aristotle1991rhetoric, cicero1942deoratore}.

Indeed, the notion that the type of genre matters greatly is strongly supported by prior linguistic research. \citet{wingrove2017suitable} demonstrated that TED Talks differ systematically from academic lectures in key textual and temporal dimensions, including lower academic lexical density, higher speech rates, and a more scripted delivery style. Based on these differences, the authors argued that TED Talks should not be treated as interchangeable with academic lectures for pedagogical purposes. Thus, applying academic evaluation criteria to TED Talks introduces a construct mismatch: the framework captures dimensions relevant to university instruction (informative speaking/docere), but only weakly related to the speaker's goals and the audience's expectations in the TED context.

\subsection{Strengths, Limitations, and Avenues for Future Research}
A major strength of this study is its scale and the use of LLMs to provide objective, multi-run evaluations of communicative quality, which bypasses the subjectivity and fatigue associated with human coding. Furthermore, the demonstration that LLM-derived clarity judgments predict real-world impact of textual products opens the door for several applications beyond TED-talk genre. Most notably, we see a large potential for using these insights to improve the clarity of classroom teaching \citep{zion2025ai}, but also areas like public science communication (e.g. about health topics, etc.).


However, like all research, several limitations deserve mention. First, our analysis was restricted to transcripts. As emphasized by  e.g. \citet{xia2021engaging}, TED Talks are inherently multimodal, relying on gestures, gaze, and visual aids. While clarity in the transcript is vital, it likely interacts with these visual cues in ways we did not measure. In several ongoing strands of research, we are examining the role of nonverbal behaviors as well as impact of clear slides and blackboard writings - all of which suggest similar mechanisms and effects as we presented here - but these topics are currently beyond the scope as technical capabilies of vision-language models are still evolving.

Future research should thus integrate LLM-based linguistic analysis with computer vision to examine the interaction between linguistic clarity and non-verbal delivery. Indeed, some work exists already exploring these directions (e.g. \citep{curtis2015effects, bernad2023multimodal, xia2021engaging}). However, very often this work is focused on elementary or molecular features (e.g. pitch of voice, discretized use of jargon-associated words), rather than the more molar, holistic concepts like clarity. Additionally, relating to the discussion about genres above, this work focused on TED Talks; investigating these dynamics in more informal or polarizing settings, like TikTok or scientific debates, would determine if the "clarity premium" holds across all genres, platforms, or digital subcultures.

\subsection{Summary and Conclusions}
Linguistic clarity is not merely a stylistic feature; it is a powerful predictor of audience impact. By leveraging AI to decode the transcripts of world-class speakers, we have shown that the clarity of an explanation is a reliable predictor of audience appreciation, significant even though the majority of TED talks are already well-composed and optimized by speakers who are keen on making an impact. The TED enterprise uses the tag line "ideas worth spreading"; our study shows that when ideas are communicated more clearly, they spread better. 

\subsection*{CRediT Author Statement}
\textbf{Roni Segal:} Conceptualization, Validation, Formal Analysis, 
Writing (Original Draft), Writing (Review \& Editing), Visualization.
\textbf{Matan Lary:} Conceptualization, Methodology, Software, Formal 
Analysis, Investigation, Data Curation, Writing (Original Draft).
\textbf{Ralf Schmaelzle:} Conceptualization, Writing (Original Draft), 
Writing (Review \& Editing).
\textbf{Yossi Ben-Zion:} Conceptualization, Methodology, Validation, 
Formal Analysis, Writing (Original Draft), Writing (Review \& Editing), 
Supervision.

\subsection*{Funding}
This research received no external funding.

\subsection*{Declaration of Competing Interest}
The authors declare no competing interests.

\subsection*{Ethics Approval}
This study used publicly available data; no ethical approval was required.

\subsection*{Generative AI}
The authors used ChatGPT and Claude to improve language clarity. 
All content was reviewed and approved by the authors.

\subsection*{Data Availability}
The data supporting the findings of this study are openly available 
on Zenodo at \url{https://doi.org/10.5281/zenodo.19391896}.

\clearpage
\appendix
\addcontentsline{toc}{section}{Supplementary Materials}

\section{Google Trends Temporal Analysis}
\label{app:google_trends}

This analysis provides supplementary visualization of the temporal context variable described in Section~\ref{subsec:google_trends}.

To provide contextual information regarding the temporal dynamics of public interest in TED, we analyzed Google Trends data for the topic “TED” between 2007 and 2013.

\begin{figure}[htbp]
    \centering
    \includegraphics[width=0.85\textwidth]{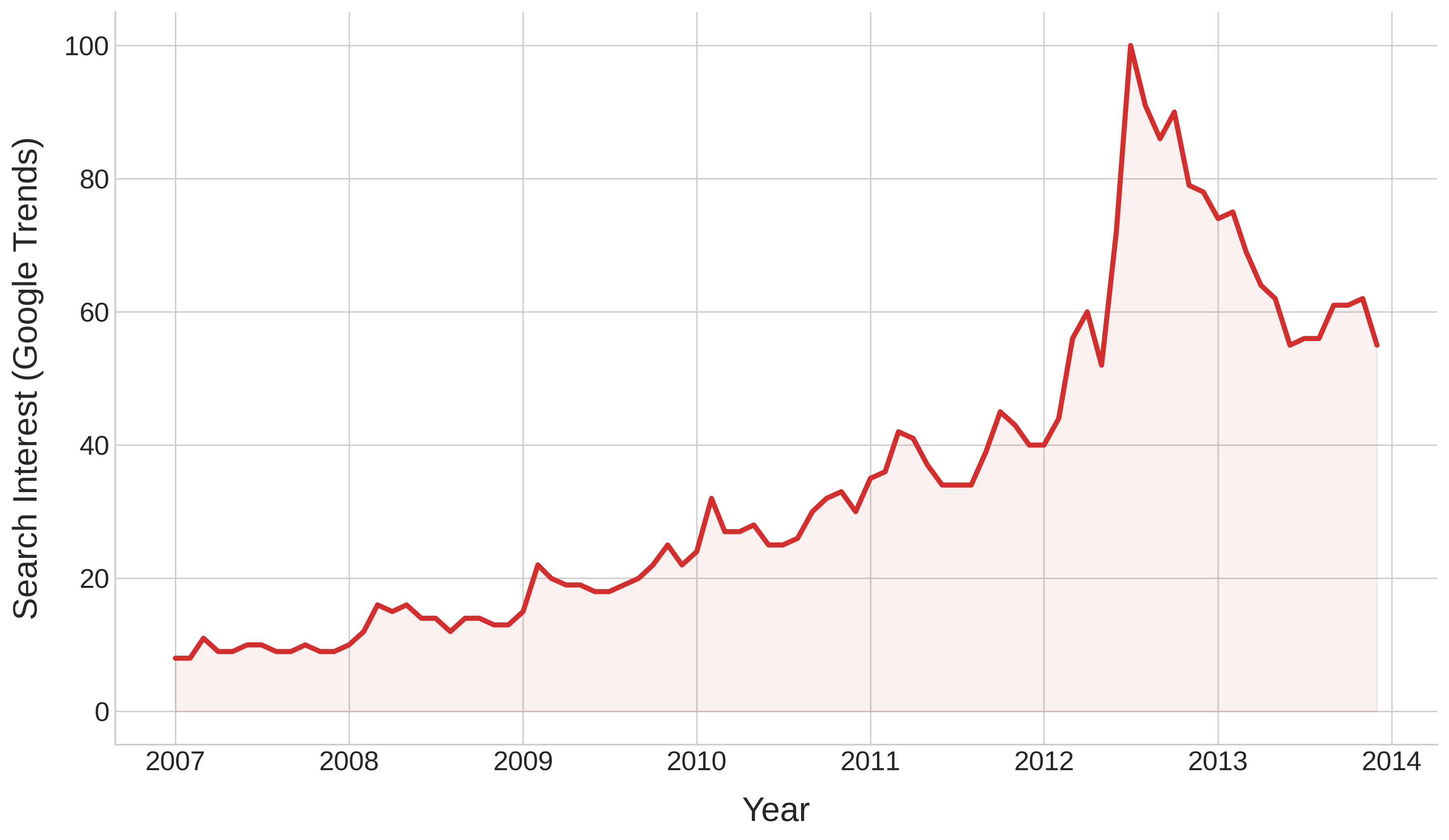}
    \caption{Global search interest for the topic \textit{“TED”} based on monthly Google Trends data (2007–2013). Values are normalized on a 0–100 scale, with 100 indicating the peak level of search activity during the period.}
    \label{fig:googletrends}
\end{figure}

As shown in Figure~\ref{fig:googletrends}, search interest increased substantially over time, reflecting the growing global visibility of TED Talks during the early expansion phase of the platform.

\section{Distribution of Scientific Classification Scores}
\label{app:scientific_hist}

This analysis corresponds to the scientific classification procedure described in Section~\ref{subsec:Sci_class}.
To assess the stability of the transcript-based scientific classification, we examined the distribution of averaged classification scores across repeated LLM evaluations.

\begin{figure}[htbp]
    \centering
    \includegraphics[width=0.85\textwidth]{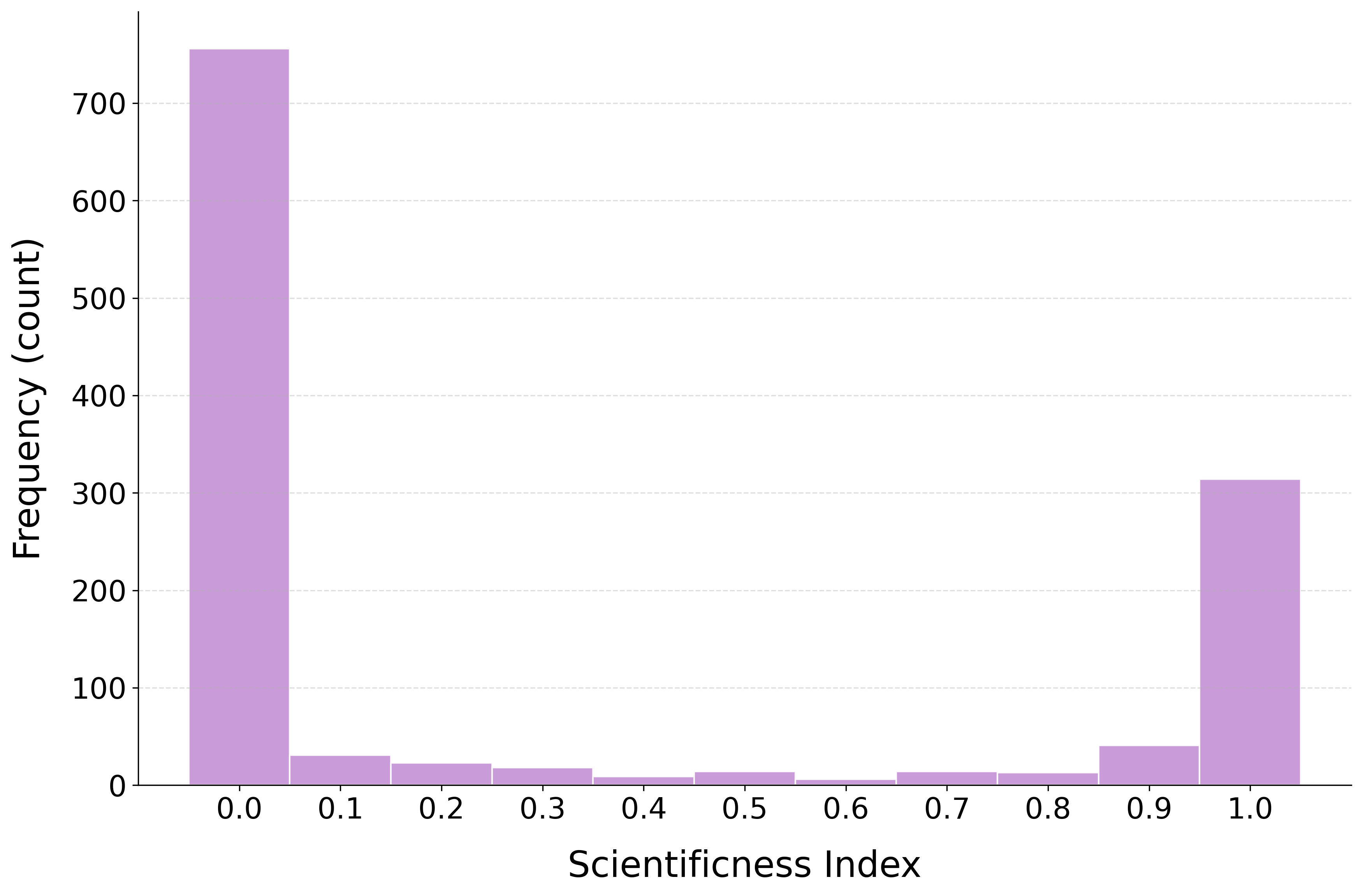}
    \caption{Distribution of scientificness scores across TED Talks (N = 1,239). Most values cluster near 0 and 1, indicating that the majority of talks were clearly classified as either scientific or non-scientific.}
    \label{fig:scientific_hist}
\end{figure}

As shown in Figure~\ref{fig:scientific_hist}, the distribution is strongly bimodal, with most values concentrated near 0 and 1. This pattern indicates high consistency across repeated LLM classifications and supports the reliability of the binary scientific labeling procedure.

\section{Academic Lecture Prompt}
\label{app:prompt}

For transparency and reproducibility, the academic lecture prompt used in the validation analysis is presented below. The TED-focused prompt, which serves as the primary evaluation method throughout this study, is presented in Section~\ref{subsec:ai_clarity}..

The academic prompt was developed for university-level physics lectures (\cite{zion2025ai}):

\begin{quote}
You will serve as a pedagogical expert in evaluating university-level teaching.\\
Your task is to assess the quality of teaching based on the following two criteria:\\[4pt]
\textit{Clarity of Explanation (1–10)}\\
\textit{Lecture Structure and Logical Flow (1–10)}\\[4pt]
Evaluate based on a transcript of a lecture where only the lecturer's speech is transcribed.\\
Provide a score between 1 and 10 for each criterion, without further explanation.\\
Your response should be in the format: \texttt{X,X} (e.g., \texttt{8,9})\\[4pt]
\{transcript\}
\end{quote}

The primary difference between the academic and TED prompts lies in the expert role definition: the academic prompt specifies ``pedagogical expert in evaluating university-level teaching'' whereas the TED prompt specifies ``expert in evaluating TED lectures''. As demonstrated in Section~\ref{subsec:Validation}, the TED prompt produced correlations with engagement metrics approximately twice as strong as the academic prompt (r = .390 vs. r = .219 for \textit{Clarity} predicting likes).

\section{Spearman Correlations}
\label{app:spearman}

Table~\ref{tab:spearman} presents the Spearman rank-order correlation coefficients among the same set of variables reported in the main correlation matrix (Table~\ref{tab:correlations}). The general pattern of associations was consistent with the Pearson results, with similar directional trends and closely aligned effect sizes across variable pairs.

\begin{table}[H]
\centering
\caption{Spearman rank-order correlations among key variables (N = 1,239).}
\label{tab:spearman}
\begin{adjustbox}{max width=\textwidth}
\begin{tabular}{lcccccc}
\toprule
\textbf{Variable} & TED\_TrendIndex & Clarity & Structure & Duration (s) & Views & Likes \\
\midrule
TED\_TrendIndex & 1 & .303** & .318** & -.305** & .186** & .316** \\
Clarity &  & 1 & .936** & -.089** & .343** & .401** \\
Structure &  &  & 1 & -.094** & .311** & .367** \\
Duration (s) &  &  &  & 1 & .103** & .090** \\
Views &  &  &  &  & 1 & .949** \\
Likes &  &  &  &  &  & 1 \\
\bottomrule
\end{tabular}
\end{adjustbox}
\begin{flushleft}
\textit{Note.} **p\textless .01 (two-tailed).
\end{flushleft}
\end{table}

\clearpage

\section{Descriptive Statistics by Content Category}
\label{app:category_descriptives}

To examine whether the predictive relationship between linguistic clarity and audience engagement varies by topical domain, we computed Pearson correlations between \textit{clarity} and both engagement measures (likes and views) separately within each of the seven content categories. As shown in Table~\ref{tab:category_correlations}, the correlations were positive across all domains and statistically significant in most categories, with comparable effect sizes across topics. The health category showed weaker and non-significant associations. This indicates that clearer explanations are generally associated with greater audience appreciation across diverse topics, although the strength of this relationship may vary modestly by domain.

\begin{table}[htbp]
\centering
\caption{Descriptive statistics and clarity--engagement correlations by content category and scientificity (N = 1,239).}
\label{tab:category_correlations}
\resizebox{\textwidth}{!}{%
\begin{tabular}{lcccccc}
\toprule
\textbf{Category} 
& \textbf{Clarity (M)} 
& \textbf{Clarity (SD)} 
& \textbf{Likes (M)} 
& \textbf{Views (M)} 
& \textbf{Clarity--Likes $r$} 
& \textbf{Clarity--Views $r$} \\
\midrule
Cosmos          & 7.88 & 0.60 & 3.28 & 5.24 & .269   & .242   \\
Entertainment   & 7.45 & 0.76 & 3.42 & 5.40 & .424** & .342** \\
Environment     & 7.85 & 0.56 & 3.13 & 5.09 & .395** & .286** \\
Health          & 7.95 & 0.52 & 3.20 & 5.15 & .170   & .112   \\
Mind            & 8.09 & 0.53 & 3.94 & 5.83 & .423** & .392** \\
Society         & 7.93 & 0.55 & 3.39 & 5.32 & .406** & .392** \\
Tech            & 7.78 & 0.57 & 3.23 & 5.22 & .393** & .327** \\
\midrule
Non--Scientific & 7.78 & 0.64 & 3.35 & 5.31 & .381** & .325** \\
Scientific      & 7.97 & 0.52 & 3.37 & 5.31 & .363** & .311** \\
\bottomrule
\end{tabular}
}
\\[1mm]
{\small Note: $^{*}p<.05$; $^{**}p<.01$.} \\
\end{table}

Beyond the stability of the clarity–engagement associations across topical categories, we additionally examined whether these relationships differ between scientific and non-scientific talks. Pearson correlations between \textit{Clarity} and \textit{Likes} were computed separately for the two groups, revealing a highly similar pattern in both: scientific talks showed $r = .363$, while non-scientific talks showed $r = .381$. This indicates that linguistic clarity predicts audience engagement robustly regardless of whether the content is scientific or not.

Taken together, the stability of these correlations across both topical categories and scientific classification supports the generalizability of clarity as a communicative cue that predicts audience response in TED-style presentations.

\subsection*{Full Hierarchical Regression With Interaction Terms (Model 4)}
\label{app:model4}

Table~\ref{tab:reg_4} presents only Step~IV of the hierarchical regression (Model~4). In this step, all Category~$\times$~Clarity interaction terms, as well as the Science~$\times$~Clarity interaction, were added to the model predicting \textit{Likes}. This specification builds directly on the baseline hierarchical models reported in~\ref{Reg1-3}, which present Steps~I--III.

As shown in the table, most interaction predictors did not reach statistical significance, with the exception of the Health~$\times$~Clarity interaction. Nevertheless, the interaction block as a whole contributed only negligible incremental variance to the model ($\Delta R^{2} = .006$).

Overall, these results indicate that allowing the clarity slope to vary across topical domains provides minimal explanatory benefit. With the exception of a modest moderation effect in the Health category, the clarity--engagement relationship appears largely invariant across TED content categories and between scientific and non-scientific talks.

\begin{table}[htbp]
\centering
\caption{Hierarchical regression predicting \textit{Likes}: Step IV interaction effects.}
\label{tab:reg_4}
\begin{tabular}{lllllll}
\toprule
 & \textbf{Predictor} & \textbf{$\beta$} & \textbf{t} & \textbf{F} & \textbf{$R^2$} & \textbf{$\Delta R^2$} \\
\midrule

\textbf{Step IV} 
 & TED\_TrendIndex & 0.230 & 8.82$^{***}$ & 30.21$^{***}$ & 0.296 & 0.006 \\
 & Duration (s) & 0.184 & 7.19$^{***}$ \\
 & Science & -0.268 & -0.63 \\
 & Cosmos (vs. Society) & 0.387 & 1.08 \\
 & Mind (vs. Society) & 0.102 & 0.25 \\
 & Tech (vs. Society) & 0.526 & 1.35 \\
 & Entertainment (vs. Society) & 0.494 & 1.61 \\
 & Health (vs. Society) & 1.173 & 2.70$^{**}$ \\
 & Environment (vs. Society) & 0.472 & 1.17 \\
 & Clarity & 0.409 & 8.58$^{***}$ \\
 & Science $\times$ Clarity & 0.291 & 0.68 \\
 & Tech $\times$ Clarity & -0.548 & -1.41 \\
 & Entertainment $\times$ Clarity & -0.337 & -1.13 \\
 & Health $\times$ Clarity & -1.268 & -2.92$^{**}$ \\
 & Environment $\times$ Clarity & -0.579 & -1.44 \\
 & Mind $\times$ Clarity & 0.079 & 0.19 \\
 & Cosmos $\times$ Clarity & -0.393 & -1.10 \\
\bottomrule
\multicolumn{7}{l}{\small Note. $^{*}p<.05$; $^{**}p<.01$; $^{***}p<.001$.} \\
\end{tabular}
\end{table}

\subsection*{Full Dataset analyses}
\label{app:full data}

The following supplementary analyses are provided for the full unfiltered dataset (N = 1,280). 
These analyses include the complete correlation matrix and the full three-step hierarchical 
regression predicting \textit{Likes}. They are reported here to demonstrate that the main results of the study remain substantively unchanged regardless of the clarity-based filtering 
applied in the primary analyses.

Table~\ref{tab:corr_full} reports the Pearson correlation coefficients among all study variables for the full dataset. 
Significance values correspond to two-tailed tests.

\begin{table}[htbp]
\centering
\caption{Pearson correlation matrix for the full unfiltered dataset (N = 1,280).}
\label{tab:corr_full}
\begin{tabular}{lcccccc}
\toprule
\textbf{Variable} & TED\_search & Clarity & Structure & Duration (s) & Views & Likes \\
\midrule
TED\_TrendIndex     & 1          & .267$^{**}$ & .257$^{**}$ & -.243$^{**}$ & .185$^{**}$ & .292$^{**}$ \\
Clarity         &             & 1           & .947$^{**}$ & -.025       & .270$^{**}$ & .330$^{**}$ \\
Structure       &             &             & 1           & -.002       & .199$^{**}$ & .262$^{**}$ \\
Duration (s)    &             &             &             & 1             & .110$^{**}$ & .109$^{**}$ \\
Views      &             &             &             &               & 1           & .964$^{**}$ \\
Likes      &             &             &             &               &             & 1           \\
\bottomrule
\multicolumn{7}{l}{\small Note: $^{*}p<.05$; $^{**}p<.01$.} \\
\end{tabular}
\end{table}

Table~\ref{tab:reg_likes_full} presents the complete three-step hierarchical regression model predicting \textit{Likes} 
for the full unfiltered dataset (N = 1,280). This model mirrors the structure reported in the main text, 
and the results remain consistent across filtered and unfiltered samples.

\begin{table}[htbp]
\centering
\caption{Hierarchical regression predicting \textit{Likes} for the full unfiltered dataset (N = 1,280).}
\label{tab:reg_likes_full}
\begin{tabular}{lllllll}
\toprule
 & \textbf{Predictor} & \textbf{$\beta$} & \textbf{t} & \textbf{F} & \textbf{$R^2$} & \textbf{$\Delta R^2$} \\
\midrule

\textbf{Step I} 
 & TED\_TrendIndex & 0.339 & 12.53$^{***}$ & 87.14$^{***}$ & 0.120 \\
 & Duration (s)    & 0.192 & 7.08$^{***}$ \\
\midrule

\textbf{Step II} 
 & TED\_TrendIndex           & 0.327 & 12.43$^{***}$ & 34.93$^{***}$ & 0.198 & 0.078 \\
 & Duration (s)              & 0.170 & 6.43$^{***}$  \\
 & Science                   & 0.069 & 2.25$^{*}$    \\
 & Cosmos (vs. Society)      & -0.015 & -0.55         \\
 & Mind (vs. Society)        & 0.194 & 6.81$^{***}$  \\
 & Tech (vs. Society)        & -0.072 & -2.41$^{*}$   \\
 & Entertainment (vs. Society) & 0.027 & 0.96         \\
 & Health (vs. Society)      & -0.110 & -3.66$^{***}$ \\
 & Environment (vs. Society) & -0.137 & -4.53$^{***}$ \\
\midrule

\textbf{Step III} 
 & TED\_TrendIndex           & 0.255 & 9.81$^{***}$  & 46.68$^{***}$ & 0.269 & 0.071 \\
 & Duration (s)              & 0.173 & 6.87$^{***}$  \\
 & Science                   & 0.029 & 1.00          \\
 & Cosmos (vs. Society)      & -0.006 & -0.23        \\
 & Mind (vs. Society)        & 0.190 & 6.97$^{***}$ \\
 & Tech (vs. Society)        & -0.036 & -1.26        \\
 & Entertainment (vs. Society) & 0.136 & 4.69$^{***}$ \\
 & Health (vs. Society)      & -0.095 & -3.29$^{**}$ \\
 & Environment (vs. Society) & -0.118 & -4.08$^{***}$ \\
 & Clarity                   & 0.301 & 11.06$^{***}$ \\
\bottomrule

\multicolumn{7}{l}{\small Note: $^{*}p<.05$; $^{**}p<.01$; $^{***}p<.001$.} \\
\end{tabular}
\end{table}

\clearpage

\bibliographystyle{model5-names}
\bibliography{ref_RS}

\end{document}